\documentclass{aa}  
\usepackage{graphicx}
\usepackage{txfonts}
\usepackage[colorlinks, citecolor=blue]{hyperref}
\begin{document}

   \title{Fundamental relation in isolated galaxies, pairs, and triplets in the local universe}
   \author{M. Argudo-Fernández\inst{1,2,3}
          \and
          S. Duarte Puertas\inst{1,2,4}
          \and
          S. Verley\inst{1,2}
          }

\institute{
Departamento de Física Teórica y del Cosmos, Edificio Mecenas, Campus Fuentenueva, Universidad de Granada, E-18071, Granada, Spain \\ \email{margudo@ugr.es}
\and 
Instituto Universitario Carlos I de F\'isica Te\'orica y Computacional, Universidad de Granada, 18071 Granada, Spain
\and
Instituto de F\'isica, Pontificia Universidad Cat\'olica de Valpara\'iso, Casilla 4059, Valpara\'iso, Chile
\and
D\'epartement de Physique, de G\'enie Physique et d'Optique, Universit\'e Laval, and Centre de Recherche en Astrophysique du Qu\'ebec (CRAQ), Qu\'ebec, QC, G1V 0A6, Canada
}

\date{Received ; accepted }

 
  \abstract
   {The correlations between star formation rate (SFR), stellar mass (M$_\star$), and gas-phase metallicity for star-forming (SF) galaxies, known as global scaling relations or fundamental relations, have been studied during the last decades to understand the evolution of galaxies. However the origin of these correlations and their scatter, which may also be related to their morphology or environment, is still a subject of debate.}
   {In this work, we establish fundamental relations, for the first time, in isolated systems in the local universe (with 0.005\,$\leq$\,z\,$\leq$\,0.080), which can give insight into the underlying physics of star-formation. We use a sample of isolated galaxies to explore whether star formation is regulated by smooth secular processes. In addition, galaxies in physically bound isolated pairs and isolated triplets may also interact with each other, where interaction itself may enhance/regulate star-formation and the distribution of gas and metals within galaxies.}
   {We made use of published emission line fluxes information from the Sloan Digital Sky Survey (SDSS) to identify SF galaxies in the SDSS-based catalogue of isolated galaxies (SIG), isolated pairs (SIP), and isolated triplets (SIT). We also use this data to derive their aperture-corrected SFR (considering two different methods) and oxygen abundance, 12 + log(O/H), using bright lines calibrations. Stellar masses for SIG, SIP, and SIT galaxies were estimated by fitting their spectral energy distribution on the five SDSS bands.}
   {The SFR results found using both methods seem to be consistent. We compared our results with a sample of SF galaxies in the SDSS. We found that, on average, at a fixed stellar mass, the SIG SF galaxies have lower SFR values than Main Sequence (MS) SF galaxies in the SDSS and central galaxies in the SIP and SIT. On average, SIG galaxies have higher 12 + log(O/H) values than galaxies in the SIP, SIT, and comparison sample. When distinguishing between central and satellite galaxies in the SIP and SIT, centrals and SIG galaxies present similar values ($\sim$8.55) while satellites have values close to M33 ($\sim$8.4). Using the $D_n$(4000) parameter as a proxy of the age of the stellar populations we found that, on average, SIG and central galaxies have higher $D_n$(4000) values than satellites and comparison galaxies.}
   {In general SIG galaxies do not present stellar starbursts produced by interactions with other galaxies and therefore their gas is consumed more slowly and at a regular pace. On the contrary, SIP and SIT galaxies present higher SFR values at fixed mass (both in central and satellite galaxies). Therefore, the effect of adding one or two companion galaxies is noticeable and produce a similar effect as the typical average environment around galaxies in the local universe. The successive interactions between the galaxies that form these pairs and triplets have enhanced the star formation. Based on our results for isolated galaxies, we propose a ground level `nurture' free SFR-M$_\star$ and gas metallicity-SFR-M$_\star$ relations for SF galaxies in the local universe.}


\titlerunning{Fundamental relation in isolated galaxies, pairs, and triplets in the local universe}
\authorrunning{M. Argudo-Fernández et al.}

   \maketitle
%

\section{Introduction}

Star formation rate (SFR), stellar mass (M$_\star$), and gas-phase metallicity (measured by the oxygen abundances, 12+log(O/H), for instance) are key physical quantities that must be considered in galaxy evolution studies \citep{2007ApJ...660L..43N,2010MNRAS.408.2115M,2012MNRAS.422..215Y,2021MNRAS.503.2340W,2022MNRAS.516.1275G,2022MNRAS.510..320F}. With a continuous inflow and collapse of cold gas, galaxies form new stars, measured by the SFR, with the subsequent increment of their stellar mass, and undergo chemical evolution, measured by the stellar or gas metallicity \citep{2004ApJ...606..271G,2009MNRAS.396L..71V,2018ApJ...868...89G,2019ApJ...884L..33L,2020ApJ...903..145L}. Stars produce heavy elements (metals) that are dispersed into the interstellar medium (ISM), contributing to the chemical enrichment of the galaxy. Gas flows (both inflows and outflows) regulate the level of metal content by diluting its gas-phase or directly expelling the enriched gas out of the galactic potential well, contributing as well to the metal enrichment of the circumgalactic medium (CGM), which extends beyond the ISM but within the virial radius \citep{2008MNRAS.385.2181F,2017ARA&A..55..389T}. Therefore, these three quantities, as physical measurements of these processes, must be somehow related. 

The correlations between these three quantities for star-forming (SF) galaxies, known as global scaling relations or fundamental relations, have been studied during the last decades. Observations have shown that there is a relation between the stellar mass and gas metallicity, the MZR relation \citep{2004ApJ...613..898T}, and between the SFR and M$_\star$, known as the main sequence for SF galaxies or Star Formation Main Sequence \citep[SFMS,][]{2004MNRAS.351.1151B, 2007ApJ...670..156D, 2010ApJ...721..193P, 2012ApJ...754L..29W, 2019MNRAS.483.3213P}. There is also a correlation of the SFR with the gas content \citep{2008ApJ...672L.107E,2013ApJ...765..140A}, especially the molecular gas, usually referred as the star formation law or Kennicutt-Schmidt (KS) law \citep{1959ApJ...129..243S,1998ApJ...498..541K}. Moreover, SFR, M$_\star$, and gas metallicity are also correlated in a fundamental plane \citep{2010MNRAS.408.2115M,2010A&A...521L..53L,2013ApJ...764..178L}, often referred to as the ``Fundamental Metallicity Relation'' (FMR or MZSFR). Hereafter we will refer to this relation as the Fundamental Relation. However the origin of these correlations and their scatter, which may also be related to their morphology or environment, is still a subject of debate \citep{2012A&A...547A..79F,2019ApJ...878L...6S,2019ApJ...882....9S, 2020MNRAS.493.1818D, 2021A&A...646A.151P, 2021MNRAS.501.1046M, 2021ApJ...918...68N, 2022A&A...666A.186D}.

As noted by \citet{2020MNRAS.491..944C}, differences can be related to the choice of the metallicity diagnostics and calibrations. However, when the measurements of all the quantities involved (especially metallicity) are performed self-consistently (and the associated observational uncertainties are properly taken into account as well), results suggest that the evolution of galaxies is regulated by smooth, secular processes and that an equilibrium condition is set between the involved physical mechanisms over cosmic time \citep{2017MNRAS.465.1384C, 2019A&A...627A..42C, 2020MNRAS.491..944C}.
In addition, \citet{2021ApJ...910..137W} recently built a theoretical framework to understand the correlation between SFR, gas mass, and metallicity, as well as their variability, which potentially uncovers the relevant physical processes of star formation at different scales, that is applicable from 100\,pc scales up to galactic scales, from individual galaxies up to the overall galaxy population, and at both low and high redshifts \citep{2018ApJ...853..179T,2021MNRAS.501.4777E,2021A&A...650A.134P}. They conclude that on galactic scales the SF and metal-enhancement is primarily regulated by the time-varying inflow rate of gas from the surrounding intergalactic medium. In this framework, a revision of these fundamental relations in galaxies evolving secularly, in an equilibrium between gas inflow, outflow, and star formation, is still missing and would provide valuable information to constraint chemical evolution models for SF galaxies \citep{2007MNRAS.375L..16C,2013ApJ...772..119L,2017ApJ...847...18Z,2022MNRAS.510..320F}. 

In this work, we establish fundamental relations, for the first time, in isolated systems in the local universe, which can give insight into the underlying physics of star formation. Compared to isolated galaxies, galaxies in isolated pairs and isolated triplets (the smallest and most isolated galaxy groups) may share a common CGM, which regulates the distribution of gas and metal \citep{1990ApJS...74..833H, 2011Sci...334..948T, 2017ARA&A..55..389T}, so that their evolution may be linked to some extent. In addition, galaxies in physically bound isolated pairs and triplets, may also interact with each other, where interaction itself enhance/regulate star-formation and the distribution of gas and metals \citep{2020MNRAS.492.6027E}. 

This work is presented as follows. In Sect.~\ref{sec:data} we present the SDSS data that we use and the samples of isolated systems. In Sect.~\ref{sec:metho}, we describe the methodology that we follow in order to obtain the SFR and gas-phase metallicity of the galaxies. Our main results are introduced in Sect.~\ref{sec:results} and the associated discussion is in Sect.~\ref{sec:discu}. Finally, a summary and the main conclusions of our work are given in Sect.~\ref{sec:conclu}. Throughout the paper, we use a cosmology with $\Omega_{\Lambda 0} = 0.7$, $\Omega_{\rm m 0} = 0.3$, and $H_0 = 70$\,km\,s$^{-1}$\,Mpc$^{-1}$.

\section{Data and samples}
\label{sec:data}

\subsection{Isolated systems in the Sloan Digital Sky Survey}
\label{subsec:samples}

The samples of isolated systems that we use in this study are based on the catalogues of isolated galaxies, isolated pairs, and isolated triplets compiled by \citet{2015A&A...578A.110A} using photometric and spectroscopic data from the Sloan Digital Sky Survey Data Release 10 \citep[SDSS-DR10,][]{2014ApJS..211...17A}. We refer to these catalogues as SIG (SDSS-based Isolated Galaxies), SIP (SDSS-based Isolated Pairs), and SIT (SDSS-based Isolated Triplets). The SDSS uses a 2.5$\,$m telescope equipped with a CCD camera to image the sky in five optical bands (u, g, r, i, z, magnitude limits: u:22.0; g:22.2; r:22.2; i:21.3; z:20.5) and two digital spectrographs \citep{2000AJ....120.1579Y, 2014ApJS..211...17A}. The full photometric catalogue for SDSS-DR10 imaging consists of 1,231,051,050 objects, of which 208,478,448 are galaxies. Similarly, the full spectroscopic catalogue for SDSS-DR10 has $\sim$3,358,200 objects, of which 1,848,851 are galaxies. 

The SDSS spectroscopic primary galaxy sample is complete in the apparent magnitude range 14.5\,$\leq$\,$\rm m_r$\, $\leq$\,17.7 \citep{2002AJ....124.1810S}. For this reason, \citet{2015A&A...578A.110A} defined a sample of primary galaxies (33,081 galaxies), with 14.5\,$\leq$\,$\rm m_r$\, $\leq$\,15.2, separated from a sample of tracers (1,607,947 galaxies), with $\rm m_r$\,$\leq$\,17.7, allowing finding neighbour galaxies up to 2.5 magnitude fainter than primary galaxies within the SDSS spectroscopic completeness. The primary sample is also limited to a redshift range 0.005\,$\leq$\,z\,$\leq$\,0.080 to discard galaxies affected by large peculiar velocities and errors or large uncertainties in their photometric measurements. An additional star-galaxy separation was applied to remove stars misclassified as galaxies by the SDSS automated pipeline. \citet{2015A&A...578A.110A} also ensured that primary galaxies have 1\,Mpc radius fields completely encircled within the photometric SDSS-DR10 footprint to avoid edge effects. 

The candidate galaxies in the SIG, and the central galaxies in the SIP and SIT,  were selected from the primary sample if there were no, one, or two neighbours within 1\,Mpc projected field radius and radial velocities differences, $\Delta\,v$, from -500 to 500\,km, s$^{-1}$ with respect to the primary galaxy, respectively. In addition, \citet{2015A&A...578A.110A} performed a visual inspection of the samples to identify any galaxy without spectroscopic information in the SDSS near primary galaxies (for example due to fiber collision), so these primary galaxies were removed from the samples. To select physically bound isolated pairs and isolated triplets, the restriction on velocity differences is up to 160\,km\,s$^{-1}$ (2$\sigma$ of the Gaussian distribution of velocity differences for satellite galaxies) within a projected distance of 450\,kpc. Under these conditions, \citet{2014A&A...564A..94A} found that $\sim$95\% of the neighbours were physically bound companions.  \citet{2015A&A...578A.110A} defined the final samples of 3,702 isolated galaxies that have been isolated for a large part of their lifetime \citep{2007A&A...474...43V}, 1,240 isolated pairs, and 315 isolated triplets.


\subsection{Selection of the star-forming galaxies}
\label{subsec:sf}

\begin{figure*}
\includegraphics[width=\textwidth]{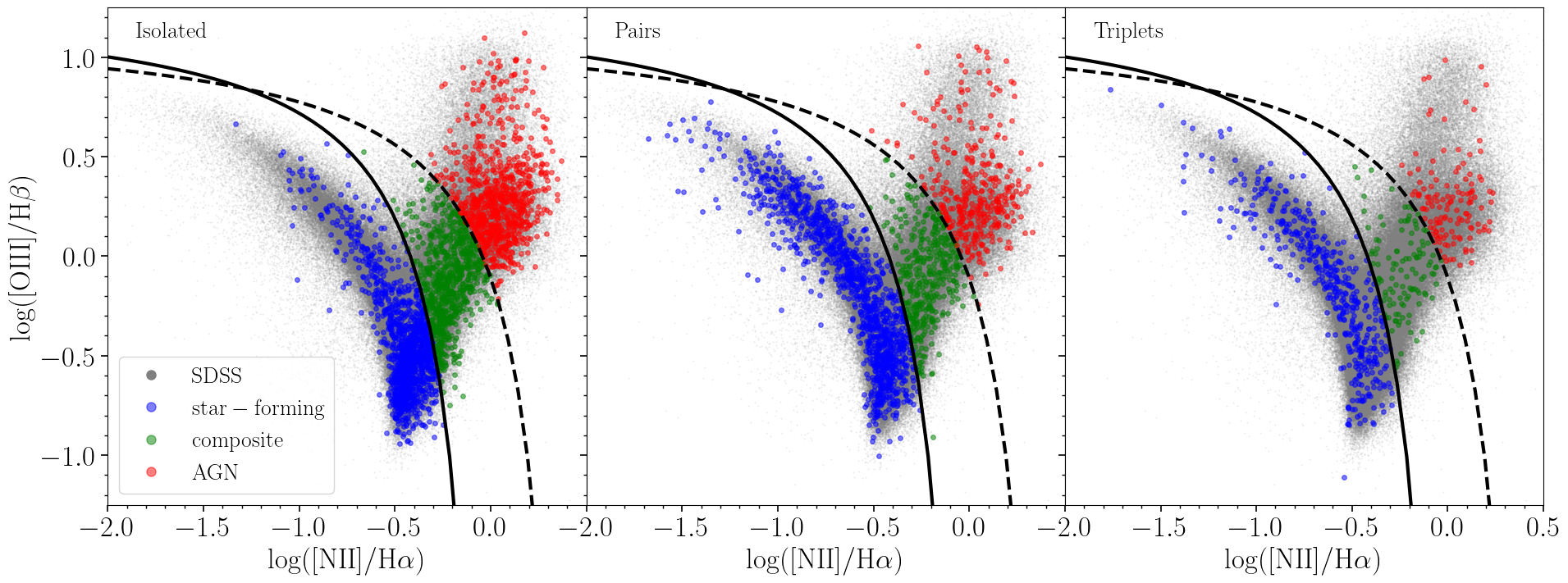}
\caption{[OIII]$\lambda$ 5007/H$\beta$ versus [NII]$\lambda$ 6583/H$\alpha$ diagnostic diagram (BPT) for isolated galaxies (left panel), galaxies in isolated pairs (middle panel), and galaxies in isolated triplets (right panel). In each panel, blue, green, and red points represent star-forming galaxies, composite galaxies, and AGN galaxies, respectively. Grey points show the position in the BPT diagram of the SDSS galaxies in the MPA-JHU catalogue. The dashed line shows the \cite{2001ApJ...556..121K} demarcation and the continuous line shows the \cite{2003MNRAS.346.1055K} curve.}
\label{Fig:BPT_diagram}
\end{figure*}

In this work we make use of the emission line fluxes provided by Max-Planck-Institut für Astrophysik and Johns Hopkins University (MPA-JHU)\footnote{Available at \href{http://www.mpa-garching.mpg.de/SDSS/}{\texttt{http://www.mpa-garching.mpg.de/SDSS/}}.} \citep{2003MNRAS.341...33K,2004MNRAS.351.1151B,2004ApJ...613..898T,2007ApJS..173..267S}. The $\sim$97\% of our sample of SIG, SIP, and SIT galaxies is included in MPA-JHU, being 3611, 2416, and 921 galaxies, respectively. Using the MPA-JHU catalogue we obtain the emission line fluxes, and their errors, of $\rm [OII]{\lambda\lambda\, 3727,\, 3729}$, $\rm [OIII]{\lambda\, 5007}$, $\rm H\beta$, $\rm [NII]{\lambda\, 6584}$, $\rm H\alpha$, and $\rm [SII]{\lambda\lambda\, 6717,\, 6731}$. We corrected all the emission line fluxes considered in this work for extinction assuming the theoretical case B recombination (the theoretical Balmer decrement, I$_{H\alpha/H\beta}$ = 2.86; T = $10^4$ K, and low-density limit $\rm n_e \sim 10^2\,cm^{-3}$; \citealt{1989agna.book.....O,1995MNRAS.272...41S}) together with the \cite{1989ApJ...345..245C} extinction curve with $\rm R_v=A_v/E(B-V)=3.1$ \citep[][]{1994ApJ...422..158O,1998ApJ...500..525S}. The selection of the subsamples of star-forming galaxies has been made on the basis of the following conditions: 

\begin{enumerate}[i)]
\item The signal-to-noise ratio\footnote{We define S/N as the ratio between the flux and the error flux.} (S/N) of the emission lines $\rm H\alpha,\ H\beta,\ [O{III}],\ and\ [N{II}]$ is $\geq$ 3. Taking into account this condition leaves subsamples of 2888 SIG, 1958 SIP, and 710 SIT galaxies.

\item We consider those galaxies that are SF, below the Kauffman demarcation in the BPT diagnostic diagram $\rm [OIII]\lambda\,5007/H\beta$ versus $\rm [NII]\lambda\,6583/H\alpha$ \citep[e.g.][]{1981PASP...93....5B,1987ApJS...63..295V,2001ApJ...556..121K,2003MNRAS.346.1055K}, as shown in Fig.~\ref{Fig:BPT_diagram}.
\end{enumerate}

After applying both conditions, we are left with  subsamples of 1282 SIG, 1209 SIP, and 460 SIT star-forming galaxies. From now on when we refer to SIG, SIP, and SIT galaxies we are referring to the subsamples of SF galaxies.

\subsection{The control and comparison sample of SDSS star-forming galaxies}
\label{subsec:comparison}

To compare the properties of star-forming galaxies in the SIG, SIP, and SIT, we use the catalogue of SDSS star-forming galaxies from \cite{2017A&A...599A..71D}, hereafter the DP17 sample. In fact, we followed the same methodology to select star-forming galaxies, and to derive SFR and oxygen abundances, as explained in Sect.~\ref{sec:metho}.
The DP17 sample is based on the MPA-JHU public catalogue, with the addition of photometric information from the SDSS-DR12 \citep{2015ApJS..219...12A}. They restricted the galaxies to a stellar mass range 10$^{8.5}$\,<\,M$_\star$/M$_\odot$\,<\,10$^{11.5}$ and redhisft 0.005\,$\leq$\,z\,$\leq$\,0.220. They applied other restrictions related to the derivation of the aperture-corrected $\rm H\alpha$ flux, see \cite{2017A&A...599A..71D} for details. Since there are not  restrictions regarding the environment, galaxies in the comparison sample can be located in any environment, mainly in denser structures (such as filaments and walls). 

\begin{figure*}
\includegraphics[width=\textwidth]{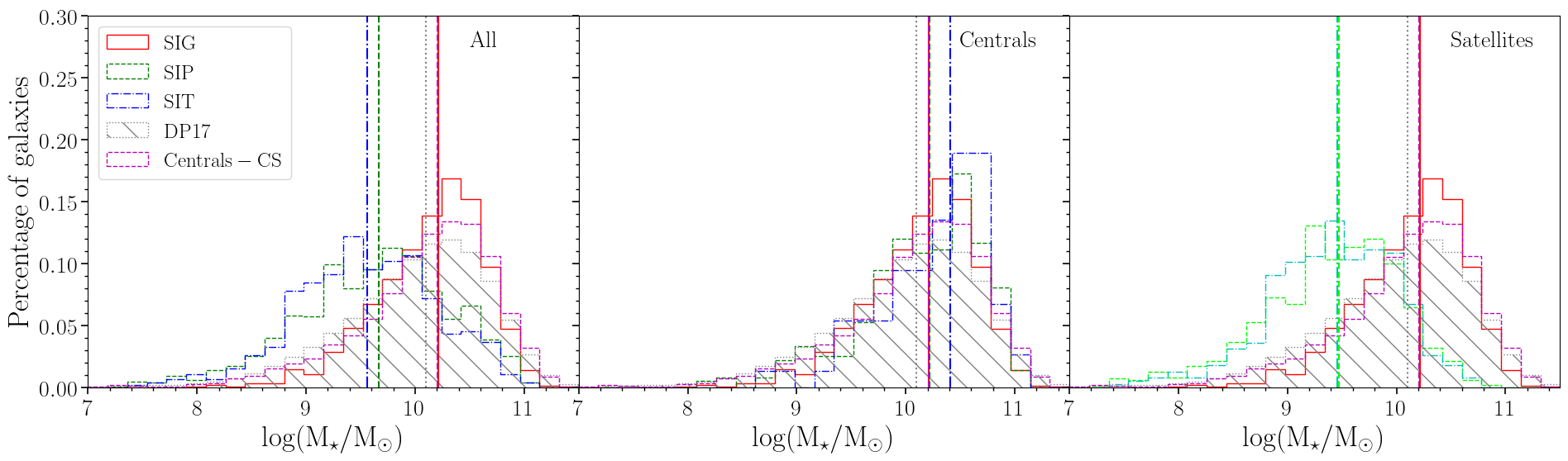} 
\caption{Distributions of the stellar mass for star-forming galaxies in the samples used in this work. From left to right: distributions for all the galaxies (including both central and satellites for isolated pairs and isolated triplets); distributions for isolated and central galaxies in isolated pairs and triplets; and distributions for satellite galaxies in isolated pairs (dash-dotted light-green distribution) and triplets (dash-dotted cyan distribution). The distributions in the first two panels follow line style and colours as indicated in the legend. Note that in the last panel we keep the distribution for isolated galaxies for reference. For reference, we also show the distribution for the DP17 sample, extended for low-mass galaxies (log(M$_\star$/M$_\odot$)\,<\,8.5) as explained in Sect.~\ref{subsec:comparison}\, and the Centrals-CS sample in all the panels. The median values for each subsample of galaxies are indicated by a vertical line, following the respective colour and line style of each distribution. The number of galaxies in each sample, as well as their corresponding median values, standard deviations, mean values and associated errors, are presented in Table~\ref{tab:meanerrprop}.}
\label{Fig:histmass}
\end{figure*}

The panels in Fig.~\ref{Fig:histmass} show the distributions of stellar mass for SIG, SIP, and SIT galaxies in comparison to the DP17 sample, extended to masses lower than 10$^{8.5}$ M$_\star$ (242385 galaxies, as shown in Table~\ref{tab:meanerrprop}). When distinguishing between central and satellite galaxies in the SIP and SIT, the central panel shows that central galaxies in the SIP and SIT have stellar mass mean values similar to the SIG and the DP17 sample, while the values for satellite galaxies are generally lower (see also Table~\ref{tab:meanerrprop}).

Note that the DP17 sample contains galaxies in a redshift range (z$\lesssim$0.22) larger than the SIG, SIP, and SIT (limited to z$<$0.08). Even if the slope of the main sequence does not significantly vary between z=0 and z=1 \citep{2019MNRAS.485.4817D}, and the mass-metallicity relation remains intact up to z\,$\sim$\,2.5 \citep{2005AIPC..761..425S,2006ApJ...644..813E,2013MNRAS.436.1130S}, we also define a control sample of SF galaxies to minimize any possible selection effects. We selected galaxies from the DP17 sample, considering the galaxies with 14.5\,$\leq$\,$\rm m_r$\,$\leq$\,15.2 and z$<$0.08 \citep[following the criteria to select primary galaxies in the SIG, SIP, and SIT as in][]{2015A&A...578A.110A}. The control sample, hereafter Centrals-CS, is composed of 5212 SF galaxies (see also Table~\ref{tab:meanerrprop}).

\begin{figure}
\includegraphics[width=\columnwidth]{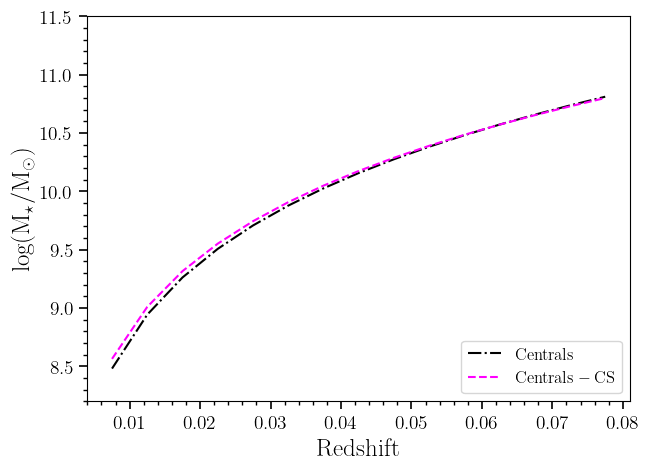} 
\caption{stellar mass completeness as a function of redshift for central galaxies (black dash-dotted line) and the control sample (magenta dashed line) within a stellar mass range $8.5\,\leq\,\rm log\,(M_{\star}/M_{\odot})\,\leq\,11.5$ (1703 central galaxies and 5086 galaxies in the control sample). The completeness has been estimated using the 95\% percentile of stellar mass at redshift bins with a width of 0.005 and applying a logarithmic fit of the form $\rm \log_{10}(M_{\star}/M_\odot)~=~A~+~B\,\log_{10}(z)$ over the redshift range.}
\label{Fig:compzmass}
\end{figure}

Figure~\ref{Fig:compzmass} shows that the Centrals-CS sample is comparable to central galaxies (i.e. SIG and central galaxies in the SIP and the SIT) in terms of completeness in stellar mass and redshift, following the methodology in \citep{2024MNRAS.532..982S} and \citep{2024MNRAS.534.1682O}. The samples are limited to a stellar mass range within $8.5\,\leq\,\rm log\,(M_{\star}/M_{\odot})\,\leq\,11.5$ to have a significant number of galaxies. Henceforth, for this study, we computed the relations between stellar mass, SFR, and oxygen abundances for central galaxies and the Centrals-CS sample, within this stellar mass range. These samples are composed of 1703 central galaxies (1275 from the SIG, 354 from the SIP, and 74 from the SIT) and 5086 galaxies in the control sample. To analyse these properties in the context of galactic conformity, we consider satellite galaxies versus central galaxies in the SIP and SIT, in all the stellar mass range (1207 galaxies from the SIP and 460 from the SIT).

\subsection{Quantification of the environment} \label{subsec:quantification_environment}

To estimate the gravitational interaction strength produced by the neighbours on the central galaxy with respect to its internal binding forces, we use the Q parameter \citep{2007A&A...472..121V,2013MNRAS.430..638S,2013A&A...560A...9A,2014A&A...564A..94A,2015A&A...578A.110A} defined as:

\begin{equation} \label{ecua_1}
Q \equiv \log \left(\sum_i {\frac{M_{i}}{M_{P}}} \left(\frac{D_P}{d_i}\right)^3\right)\quad.
\end{equation} 

\noindent The subscripts $p$ and $i$ are for primary and neighbour galaxies, respectively, where $M$ is the stellar mass of the galaxy, $D_P$ is the diameter of the primary galaxy, and $d_i$ is the projected physical distance of the i$^{\rm th}$ neighbour to the primary galaxy. Stellar masses were estimated using the five optical bands (ugriz) with the routine kcorrect \citep{2007AJ....133..734B} with a \citep{2001MNRAS.322..231K} universal initial
mass function (IMF). 

We use this tidal strength parameter, provided by \citet{2015A&A...578A.110A}, to consider the effects of the local environment on primary galaxies, henceforth centrals, by the companion galaxies, henceforth satellites, in isolated pairs (Q$_{\rm pair}$) and isolated triplets (Q$_{\rm trip}$).


\section{Methodology}
\label{sec:metho}

\subsection{SFR derivation}
\label{subsec:sfr}

It is known that the SDSS fibres (3$^{\prime\prime}$ diameter) cover a limited portion of the galaxies in the local universe (z $<$ 0.1), which implies that only a limited amount of the H$\alpha$ emission of each galaxy is measured. Using aperture corrections is necessary when we derive an extensive quantity, i.e. the value of this quantity depends on the portion of the galaxy considered. This occurs when we derive the total SFR of the SDSS star-forming galaxies at low redshift. The difference between the total SFR and the SFR in the fibre is, on average, $\sim$0.65 dex \citep{2017A&A...599A..71D}.

The SFR is derived from the H$\alpha$ luminosity and the conversion factor $\rm \eta_{H\alpha}$ \citep{2004MNRAS.351.1151B}, where $\rm SFR~=~L(H\alpha)/\eta_{H\alpha}$. The parameter $\rm \eta_{H\alpha}$ varies with the physical properties of the galaxy, total stellar mass, and metallicity \citep[see][]{2002MNRAS.330..876C,2003A&A...410...83H,2004MNRAS.351.1151B}.\footnote{Eq. 6 in \cite{2017A&A...599A..71D} parameterised the $\rm \eta_{H\alpha}$ parameter, see also Fig. 7 in \cite{2004MNRAS.351.1151B} for more details.}

In this work, SIG, SIP, and SIT galaxies have been aperture corrected using the two different recipes described in \cite{2017A&A...599A..71D}: i) the empirical \cite{2013A&A...553L...7I,2016ApJ...826...71I} aperture corrections from the CALIFA survey; and ii) the refined method of \cite{2003ApJ...599..971H} as used in \cite{2017A&A...599A..71D}. \cite{2003ApJ...599..971H} assumed that the $\rm H\alpha$ emission-line flux can be traced across the whole galaxy by the r-band emission. They derived the total $\rm H\alpha$ flux as a function of the difference between the total petrosian r-band magnitude ($\rm r_{petro}$) and the corresponding magnitude within the SDSS fibre ($\rm r_{fibre}$) as: $\rm F_{H\alpha}^{corr}=F_{H\alpha}^{0}\times 10^{-0.4(r_{petro}-r_{fibre})}$ \citep[see also][]{2013MNRAS.432.1217P}. \cite{2017A&A...599A..71D} refined this methodology taking into account that the parameter $\rm \eta_{H\alpha}$ is not a constant value and recalculated the SFR for all the SDSS star-forming galaxies in the MPA-JHU catalogue. The SFRs derived by using both methods are consistent \citep[see ][for more details]{2017A&A...599A..71D}. 

In addition to SFR, we are also using $D_n$(4000) from the MPA-JHU catalogue for this work. In particular, we use the $D_n$(4000) that corresponds to the narrow definition of the 4000$\AA$ break strength from \citet{1999ApJ...527...54B}. We use this parameter as the stellar population age indicator, which is small for galaxies with
younger stellar populations, and large for older stellar populations. We use the divisory value at $D_n$(4000)~=~1.67 following \citet{2006MNRAS.370..721M}.

\subsection{O/H derivation}
\label{subsec:oh}

We derive the oxygen abundance, 12 + log(O/H), using bright lines calibrations (i.e. [OII], [OIII], H$\beta$, [NII], H$\alpha$, [SII]) due to the fact that the faint temperature sensitive lines are not available in our sample of galaxies. \citet{2021A&A...645A..57D, 2022A&A...666A.186D} performed a comparison between different empirical methods \citep{2009MNRAS.398..949P,2016MNRAS.457.3678P,2017MNRAS.465.1384C} and the HII-CHI-mistry code \citep{2014MNRAS.441.2663P} to derive the oxygen abundance. As can be seen in Fig. 1 of \citet{2022A&A...666A.186D}, a qualitative check was made by comparing the stellar mass -- metallicity relation (MZR) for the Milky Way and 10 galaxies with: i) precise oxygen abundance values derived by the direct method from available electron temperature measurements with a sample of $\sim$200,000 SDSS star-forming galaxies; and ii) with the oxygen abundance derived using the R-calibrator of \cite{2016MNRAS.457.3678P}. The figure shows that these 11 galaxies are contained in the footprint of SDSS star-forming galaxies. For this reason, in this work we have considered the empirical calibrations proposed by \cite{2016MNRAS.457.3678P}.

We derive the oxygen abundance using the empirical calibration S of \cite{2016MNRAS.457.3678P}\footnote{The empirical calibrators are calibrated against direct derivations of abundances that are taken into account electron temperature measurements.}. This calibrator uses the [SII] emission line instead of [OII]. We use the calibrator S because for those SDSS galaxies with z\,$<$\,0.020, the emission line $\rm [OII]{\lambda, 3727, 3729}$ is not covered in the spectra. For the $\sim$200\,000 SDSS star-forming galaxies sample, the average difference between the oxygen abundance values derived with the S and R calibrators from \cite{2016MNRAS.457.3678P} is $\sim$0.03 dex, a value much lower than the average error of the calibrators.

For the derivation of the oxygen abundance we have only considered those star-forming galaxies whose S/N for $\rm [SII]{\lambda\lambda\, 6717,\, 6731}$ is $\geq$ 3. After taking into account this condition we are left with 1269 SIG, 1191 SIP, and 446 SIT galaxies.

\begin{table*}
\begin{center}
\caption{\label{table:data}Compiled data for our sample of star-forming galaxies.}
\begin{tabular}{cccccc}
\hline
\hline \\[-2ex]
 (1) & (2) & (3) & (4) & (5) & (6)\\
specObjID & $\rm \log({M_{\star}})$ & log(SFR) & 12+log(O/H) & $D_n$(4000) & Catalogue \\
 & [$\rm M_{\odot}$] & [$\rm M_{\odot}\,yr^{-1}$] & & & \\
\hline \\
995351907448940544 & 9.95$\pm$0.10 & 0.41$\pm$0.03 & 8.49$\pm$0.01 & 1.20$\pm$0.01 & SIG\\
891794679312443392 & 10.50$\pm$0.10 & 0.34$\pm$0.09 & 8.60$\pm$0.02 & 1.58$\pm$0.03 & SIG\\
464046166120622080 & 9.68$\pm$0.10 & 0.41$\pm$0.02 & 8.33$\pm$0.00 & 1.10$\pm$0.01 & SIG\\
2052605755323869184 & 10.56$\pm$0.10 & 0.69$\pm$0.04 & 8.59$\pm$0.01 & 1.34$\pm$0.02 & SIG\\
2848665917645154304 & 10.53$\pm$0.10 & $-$0.06$\pm$0.07 & 8.60$\pm$0.02 & 1.46$\pm$0.01 & SIG\\
2568209360727599104 & 10.61$\pm$0.10 & 0.77$\pm$0.03 & 8.61$\pm$0.01 & 1.28$\pm$0.01 & SIG\\
1148505361804716032 & 7.75$\pm$0.10 & $-$1.51$\pm$0.03 & 8.00$\pm$0.03 & 1.06$\pm$0.04 & SIP\\
2046971318942853120 & 7.84$\pm$0.10 & $-$2.05$\pm$0.01 & 8.01$\pm$0.06 & 1.25$\pm$0.06 & SIP\\
369358979404425216 & 10.03$\pm$0.10 & $-$0.40$\pm$0.10 & 8.59$\pm$0.02 & 1.2$7\pm$0.02 & SIP\\
498809974038751232 & 9.79$\pm$0.10 & $-$0.43$\pm$0.02 & 8.63$\pm$0.00 & 1.37$\pm$0.01 & SIT\\
... & ... & ... & ... & ... & ...\\
\hline
\end{tabular}
\tablefoot{The list of galaxies for each sample is randomly sorted. The full table is available in electronic format at the CDS. The columns correspond to: (1) SDSS spectroscopic name; (2) $\rm \log({M_{\star}})$, stellar mass from \citet{2015A&A...578A.110A} in $\rm M_{\odot}$; (3) log(SFR), aperture corrected star formation rate as described in Sect.~\ref{subsec:sfr}, in $\rm M_{\odot}\,yr^{-1}$; (4) 12+log(O/H), oxygen abundance as described in Sect.~\ref{subsec:oh}; (5) $D_n$(4000), narrow definition of the 4000$\AA$ break strength in \citet{1999ApJ...527...54B} from the MPA-JHU catalogue; and (6) indicates if the galaxy is coming from the SIG, SIP, or SIT catalogues.}
\end{center}
\end{table*}

\section{Results}
\label{sec:results}

In this work we have used aperture corrected SFR considering two different methods, using the empirical aperture corrections of \cite{2013A&A...553L...7I,2016ApJ...826...71I} and by using the refined method of \cite{2003ApJ...599..971H} as applied in \cite{2017A&A...599A..71D}. The SFR results found in both cases are consistent, with the standard deviation of the difference is $\sim$0.13 dex. Throughout the next sections, we will show the results using the second method. As a result, we have compiled Table~\ref{table:data} where we show a sample of the online table of our star-forming galaxies. In column 1 we show the SDSS spectroscopic name. Column 2 shows the stellar mass. Columns 3, 4, and 5 show the SFR, the oxygen abundance, and the $D_n$(4000). Table~\ref{table:data} also shows the uncertainties for each property considered\footnote{The uncertainties for the $D_n$(4000) are taken from MPA-JHU.}. The propagation of the uncertainties for the derivation of the SFR and the oxygen abundance have been estimated using the python package uncertainties\footnote{Uncertainties: a Python package for calculations with uncertainties, Eric O. LEBIGOT, \url{http://pythonhosted.org/uncertainties/}}. Since there is no stellar mass estimation in the MPA-JHU for all the SIG, SIP, and SIT SF galaxies, we use stellar masses in \cite{2015A&A...578A.110A}, estimated using k-correct, via a PCA-like technique based on SPS models, which do not have associated errors, as explained in \cite{2007AJ....133..734B}. Estimates of the uncertainty from the SPS model in the order of 0.10\,dex \citep{2010ApJ...717..379B}, which is also in agreement with the mean error of stellar masses in MPA (0.12\,dex) and in the comparison sample (0.09\,dex). We therefore adopt a value of 0.10\,dex for the error of the stellar masses, which is also implemented for the estimation of the uncertainties of the coefficients in our fits.

\begin{figure*}
\includegraphics[width=\textwidth]{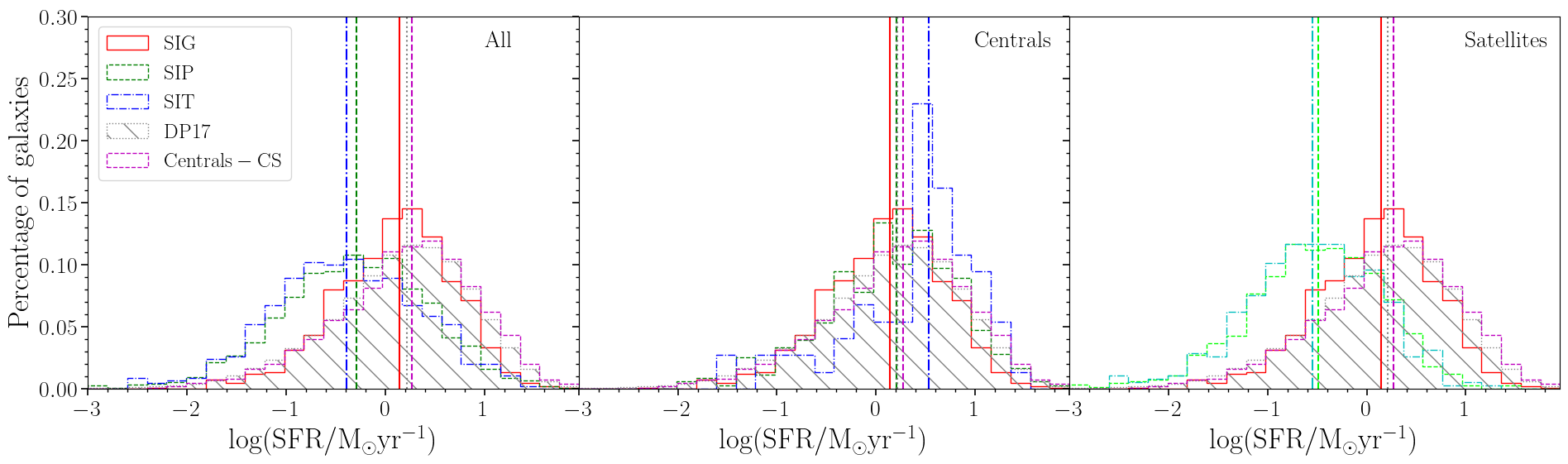} \\
\includegraphics[width=\textwidth]{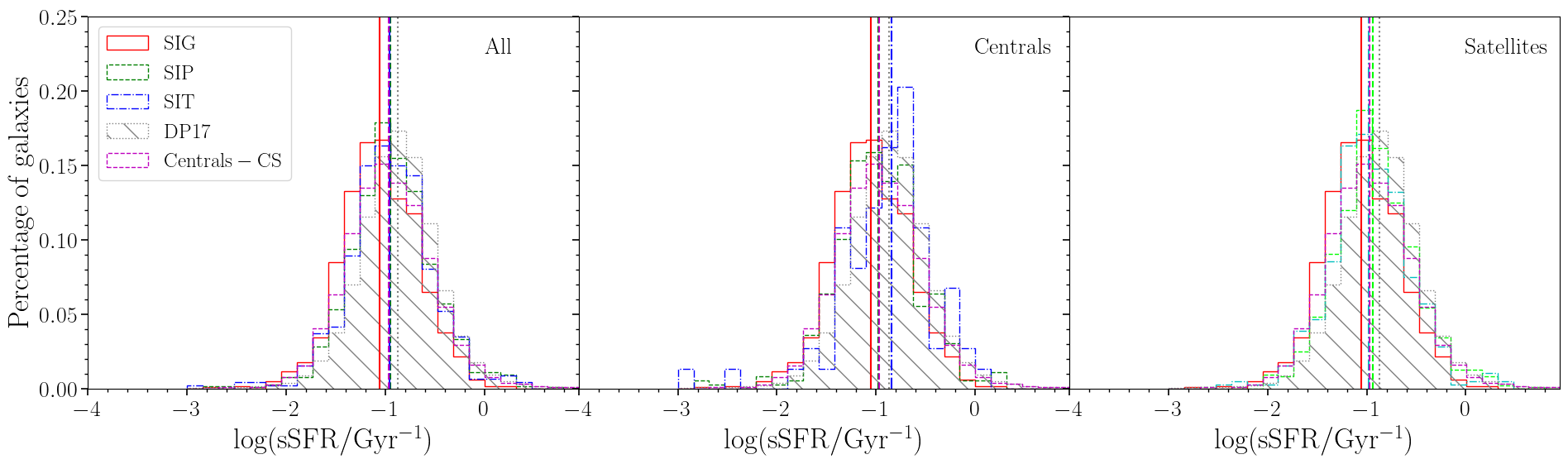} \\
\includegraphics[width=\textwidth]{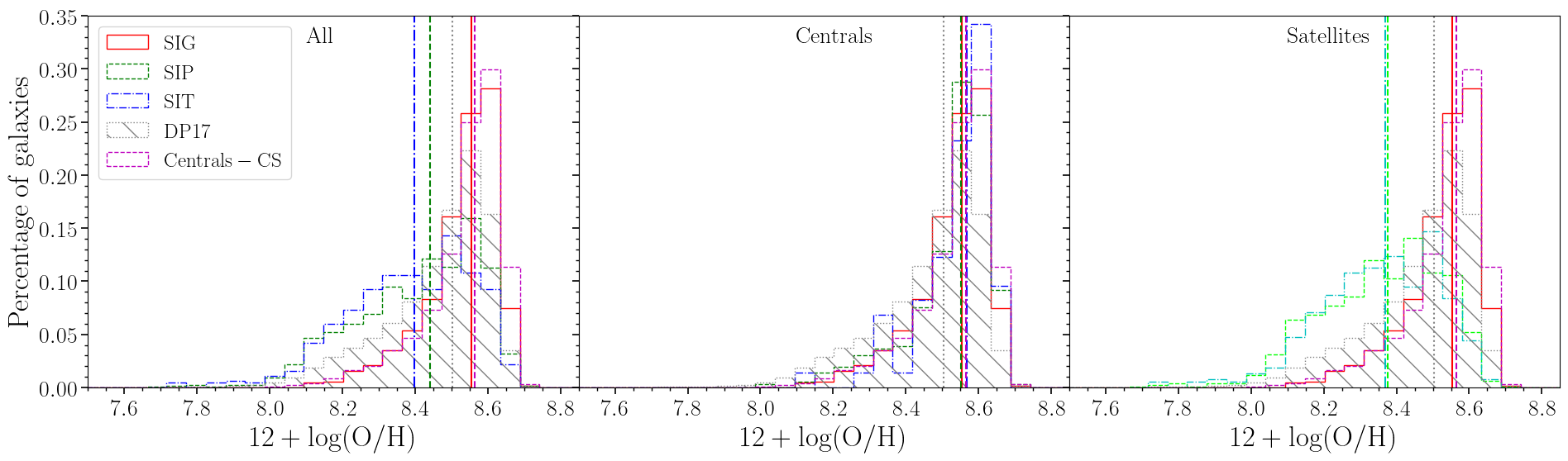} \\
\includegraphics[width=\textwidth]{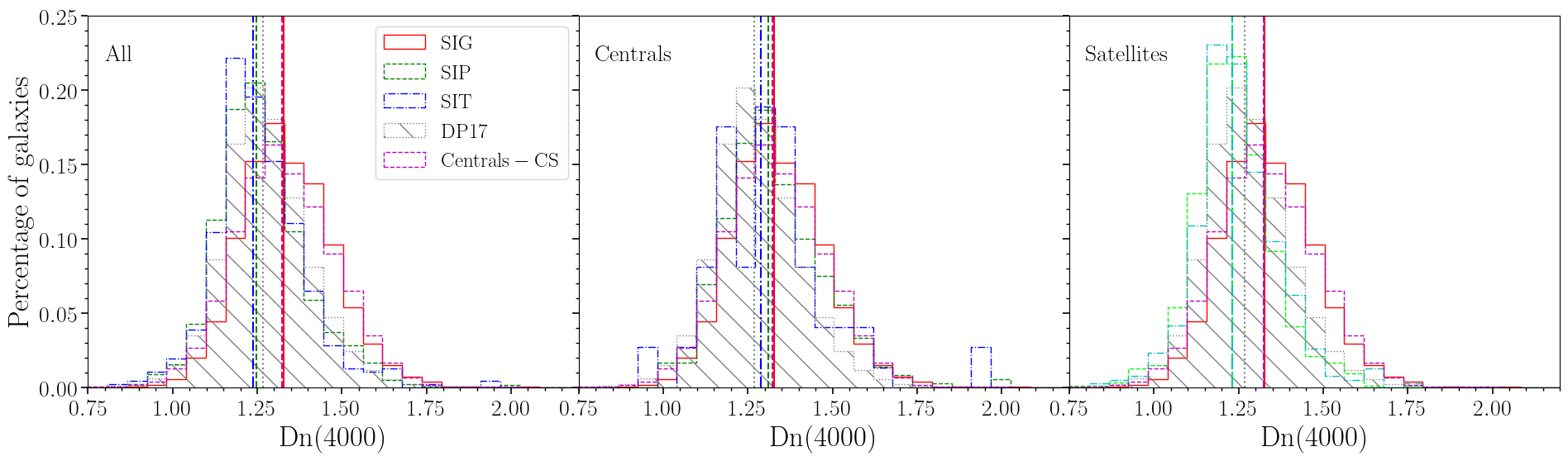}
\caption{Similar to Fig.~\ref{Fig:histmass}, distributions of the SFR, sSFR, oxygen abundance, and $D_n$(4000), from upper to lower panels, for star-forming galaxies in the samples used in this work. The median values for each subsample of galaxies are indicated by a vertical line, following the respective colour and line style of each distribution. The number of galaxies in each sample, as well as their corresponding median values, standard deviations, mean values and associated errors, are presented in Table~\ref{tab:meanerrprop}.}
\label{Fig:hists}
\end{figure*}

The distribution of the physical parameters, including the specific star formation rate (sSFR\,=\,SFR/M$_\star$), for the galaxies in our samples is shown in Fig.~\ref{Fig:hists}. The median values for each subsample of galaxies are marked in each panel with dashed lines (red for SIG, green for SIP, blue for SIT, and grey for DP17 galaxies and magenta for Centrals-CS galaxies, where Table~\ref{tab:meanerrprop} shows the median and mean values and corresponding errors. The median values for SIP and SIT galaxies are lower than SIG and DP17 galaxies (about 0.5\,dex), which is mainly caused by satellite galaxies, which generally have lower SFR values. These differences disappear when considering the sSFR (as shown in the second row of panels in Fig.~\ref{Fig:hists}).


We present our results as follows: first, in Sec.~\ref{sec:sfrmass_results}, we show the SFR -- M$_\star$ relation of SIG and central galaxies in the SIP and SIT, in the stellar mass range $8.5\,\leq\,\rm log\,(M_{\star}/M_{\odot})\,\leq\,11.5$. Second, in Sec.~\ref{subsec:ohmass_results}, we present the 12+log(O/H) -- M$_\star$ relation, while in Sec.~\ref{subsec:fundamental_results} we show the fundamental relation for the same samples of galaxies. In Sec.~\ref{sec:local_result} we explore the dependence of the local environment on the SFR and 12+log(O/H) for galaxies in isolated pairs and triplets and if there is any dependence in terms of galactic conformity. 

\begin{table*}
\centering
\caption{\label{tab:meanerrprop}Mean and median values of the distributions of physical properties for SF galaxies in this study.}
\begin{tabular}{lcccccccc}
\hline
\hline   
(1) & (2) & (3) & (4) & (5) & (6) & (7) & (8) & (9)\\
sample & subsample & $\rm \log(M_\star/M_\odot)$ & $\rm \log(SFR)$ & $\rm \log(sSFR)$ & 12+log(O/H) & $D_n$(4000) & N & N$_{\rm lim}$ \\
 &  & & [$\rm M_{\odot}\,yr^{-1}$] & [$\rm Gyr^{-1}$] &  &  & Galaxies & Galaxies \\
\hline 
DP17 & -- & 10.00$\pm$0.09 & 0.13$\pm$0.07 & $-$0.87$\pm$0.18 & 8.46$\pm$0.02 & 1.28$\pm$0.03 &  242933 & 220192 \\
 & & 10.10$\pm$0.68 &  0.22$\pm$0.74 & $-$0.87$\pm$0.46 & 8.50$\pm$0.15 & 1.27$\pm$0.14 & & \\ 
 \hline
Centrals-CS & -- & 10.08$\pm$0.10 & 0.15$\pm$0.05 & $-$0.93$\pm$0.11 & 8.53$\pm$0.01 & 1.33$\pm$0.02  & 5212 & 5086 \\
 & & 10.21$\pm$0.68 & 0.22$\pm$0.67 & $-$0.97$\pm$0.53 & 8.56$\pm$0.11 & 1.32$\pm$0.16 \\  
\hline
SIG & -- & 10.14$\pm$0.10 & 0.10$\pm$0.05 & $-$1.05$\pm$0.12 & 8.53$\pm$0.01 & 1.34$\pm$0.02 &  1282 & 1275 \\
 & & 10.21$\pm$0.49 & 0.15$\pm$0.59 & $-$1.05$\pm$0.39 & 8.55$\pm$0.10 & 1.33$\pm$0.15 &  & \\
\hline
SIP & -- & 9.61$\pm$0.10 & $-$0.31$\pm$0.06 & $-$0.93$\pm$0.13 & 8.41$\pm$0.02 & 1.25$\pm$0.03 & 1208 & -- \\
 & & 9.67$\pm$0.71 & $-$0.29$\pm$0.77 & $-$0.95$\pm$0.48 & 8.44$\pm$0.17 & 1.25$\pm$0.14 & &  \\
 & central & 10.14$\pm$0.10 & 0.16$\pm$0.05 & $-$0.98$\pm$0.11 & 8.52$\pm$0.01 & 1.33$\pm$0.02 &  359 & 354 \\
 & & 10.22$\pm$0.56 & 0.22$\pm$0.67 & $-$0.98$\pm$0.42 & 8.55$\pm$0.11 & 1.31$\pm$0.15 &  & \\
 & satellite & 9.39$\pm$0.10 & $-$0.51$\pm$0.07 & $-$0.90$\pm$0.13 & 8.36$\pm$0.02 & 1.24$\pm$0.03 &  849 & -- \\
 & & 9.47$\pm$0.64 & $-$0.49$\pm$0.72 & $-$0.94$\pm$0.50 & 8.38$\pm$0.16 & 1.23$\pm$0.12 &  & \\
 \hline
 SIT & -- & 9.55$\pm$0.10 & $-$0.40$\pm$0.06 & $-$0.95$\pm$0.12 & 8.38$\pm$0.02 & 1.26$\pm$0.03 &  460 & -- \\
  & & 9.56$\pm$0.66 & $-$0.39$\pm$0.74 & $-$0.95$\pm$0.44 & 8.40$\pm$0.17 & 1.24$\pm$0.14 &  & \\
 & central & 10.27$\pm$0.10 & 0.38$\pm$0.04 & $-$0.90$\pm$0.11 & 8.54$\pm$0.01 & 1.32$\pm$0.01 &  74 & 74 \\
  & & 10.41$\pm$0.50 & 0.53$\pm$0.66 & $-$0.84$\pm$0.48 & 8.57$\pm$0.10 & 1.29$\pm$0.17 & &  \\
 & satellite & 9.41$\pm$0.10 & $-$0.55$\pm$0.07 & $-$0.96$\pm$0.13 & 8.35$\pm$0.02 & 1.24$\pm$0.03 &  386 & -- \\
 & & 9.46$\pm$0.59 & $-$0.54$\pm$0.66 & $-$0.98$\pm$0.44 & 8.37$\pm$0.16 & 1.23$\pm$0.13 & &  \\
 \hline
\end{tabular}
\tablefoot{Mean values with associated errors (first row) and median values with standard deviation (second row) of the distribution of the $\rm \log(M_\star/M_\odot)$, $\rm \log(SFR)$, $\rm \log(sSFR)$, 12+log(O/H), and $D_n$(4000) parameter for star-forming galaxies in the samples considered in this work. The last columns show the total number of SF galaxies in each sample, and when limiting to the stellar mass range $8.5\,\leq\,\rm log\,(M_{\star}/M_{\odot})\,\leq\,11.5$ for central galaxies, respectively.}
\end{table*}

\subsection{SFR as a function of the M\texorpdfstring{$_\star$}{star}}
\label{sec:sfrmass_results}

\begin{figure*}
\includegraphics[width=\columnwidth]{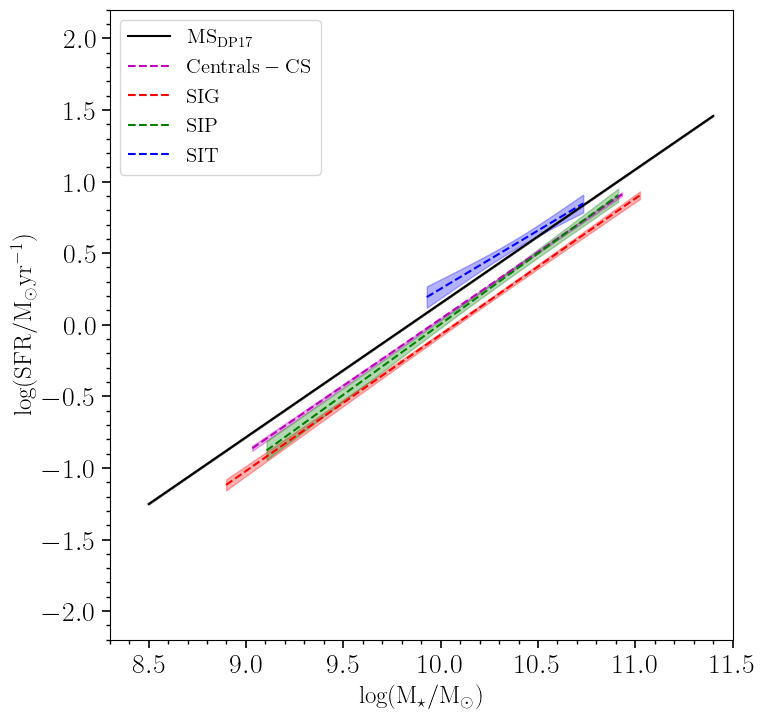}
\includegraphics[width=0.97\columnwidth]{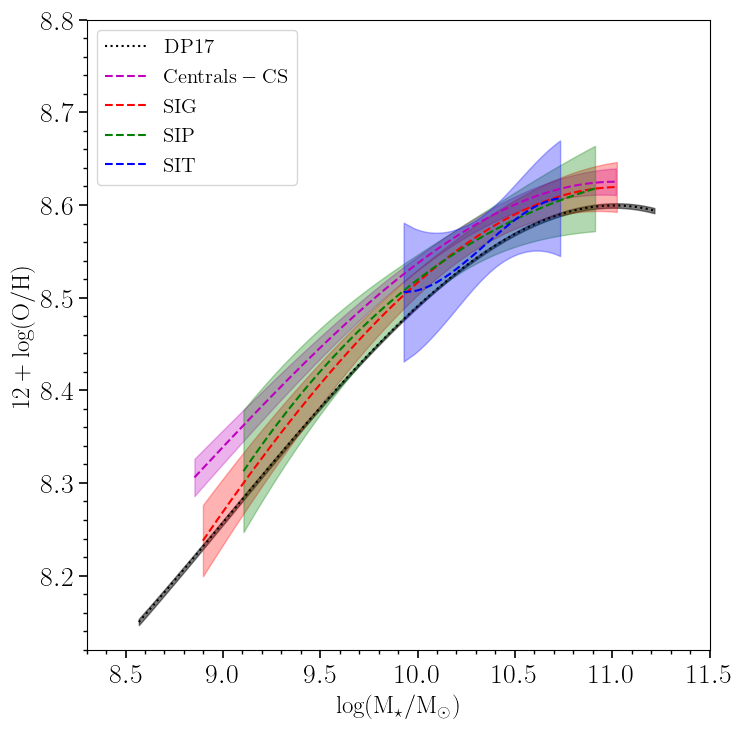}
\caption{Relation between the SFR and M$_\star$ (left panel) and the 12+log(O/H) and M$_\star$ (right panel) for central star forming galaxies with $8.5\,\leq\,\rm log\,(M_{\star}/M_{\odot})\,\leq\,11.5$. The relations are presented as the first and third order fit (respectively) to the running median in both axes. The relations for SIG galaxies (red dashed line) and central galaxies in the SIP and SIT (in green and blue dashed lines, respectively) are represented with their corresponding 1\,$\sigma$ error. For comparison, we also show the relations for star forming galaxies in DP17 and the control sample (black dotted line and grey dashed line, respectively). The number of galaxies considered in bin for each sample, as well as the parameters of the fits with their corresponding uncertainties,  are presented in Tab.~\ref{tab:unMS} and Tab.~\ref{tab:unOH}, respectively.}
\label{Fig:sfr_m}
\end{figure*}

In the left panel of Fig.~\ref{Fig:sfr_m} we show the SFR -- M$_\star$ relation for star-forming galaxies in the SIG, and central galaxies in the SIP and SIT, compared to the  Centrals-CS sample (control sample) and the relation in \citet{2017A&A...599A..71D} for DP17 galaxies (comparison sample), in the stellar mass range $8.5\,\leq\,\rm log\,(M_{\star}/M_{\odot})\,\leq\,11.5$. 
The MS for each sample has been obtained as the linear fit to the running median of the data in the form of $\rm \log(SFR)~=~a\,\log(M_\star) \, -\, b$. The parameters of the fits, with their corresponding uncertainties, and the number of bins considered to calculate the running median, are shown in Table~\ref{tab:unMS}. The slope of the SFR -- M$_\star$ relation is comparable in all samples, which is expected since the slope of the main sequence does not vary significantly between z=0 and z=1 \citep{2019MNRAS.485.4817D}. 
However, the effect of the addition of one companion galaxy is noticeable in the SIG, SIP and SIT samples, with higher SFR values at fixed stellar mass. 
The mean $\rm \Delta\,MS_{SDSS}$ for SIG galaxies is $-$0.21\,dex, slightly lower than for SIP and SIT galaxies ($-$0.13\,dex and 0.08\,dex, respectively).

\begin{table}[]
\centering
\caption{\label{tab:unMS}Parameters of the MS for SF galaxies in this study.}
\begin{tabular}{lccc}
\hline
\hline \\[-2ex]
 (1) & (2) & (3) & (4) \\
sample & bins & a & b \\
 & & [log(yr$^{-1}$)] & [log($\rm M_{\odot}\,yr^{-1}$)]  \\
\hline
DP17\tablefootmark{a} & 2000 & 0.935 $\pm$ 0.001 & $-$9.208 $\pm$ 0.001 \\
Centrals-CS & 300 & 0.934 $\pm$ 0.001 & $-$9.302 $\pm$ 0.008 \\
SIG & 30 & 0.952 $\pm$ 0.005 & $-$9.586 $\pm$ 0.052 \\
SIP & 30 & 0.987 $\pm$ 0.010 & $-$9.867 $\pm$ 0.103 \\
SIT & 30 & 0.814 $\pm$ 0.049 & $-$7.887 $\pm$ 0.503 \\
\hline
\end{tabular}
\tablefoot{Parameters of the linear fit to the running median for all SF galaxies in the in the DP17, Centrals-CS, and SIG samples, and central SF galaxies in the SIP and SIT samples, used to obtain their corresponding MS, as shown in the left panel of Fig.~\ref{Fig:sfr_m}. The uncertainties of the parameters have been estimated considering the propagation of errors when performing the linear least squares fit.} \tablefoottext{a}{Parameters from \citet{2017A&A...599A..71D}.}
\end{table} 


\subsection{Oxygen abundance as a function of the M\texorpdfstring{$_\star$}{star}}
\label{subsec:ohmass_results}

In the right panel of Fig.~\ref{Fig:sfr_m}, we show the MZR relation as the oxygen abundance, 12+log(O/H), versus the total stellar mass for star-forming galaxies in the SIG, and central galaxies in the SIP and SIT, compared to galaxies in the DP17 and Centrals-CS samples, in the stellar mass range $8.5\,\leq\,\rm log\,(M_{\star}/M_{\odot})\,\leq\,11.5$.  
The MZR for each sample has been obtained as the third order fit to the running median of the data in the form of $\rm 12\,+\,\log(O/H)\,=\,a\,\log(M_\star)^3~+~b\,\log(M_\star)^2~+~ c\,\log(M_\star)~+~d$. 
The MZR relation in all samples is comparable within 1\,$\sigma$.
The parameters of the fits, with their corresponding uncertainties, and the number of bins considered to calculate the running median, are shown in Table~\ref{tab:unOH}.

\begin{table*}[]
\centering
\caption{\label{tab:unOH}Parameters of the MZR for SF galaxies in this study.}
\begin{tabular}{lccccc}
\hline
\hline \\[-2ex]
 (1) & (2) & (3) & (4) & (5) & (6) \\
sample & bins & a & b & c & d\\
 & & [$\rm \log(M_{\odot})^{-3}$] & [$\rm \log(M_{\odot})^{-2}$] & [$\rm \log(M_{\odot})^{-1}$] &   \\
\hline
DP17 & 2000 & $-$0.0215 $\pm$ 0.0002 & 0.584 $\pm$ 0.005 & $-$5.02 $\pm$ 0.05 & 21.8 $\pm$ 0.2 \\
Centrals-CS & 300 & $-$0.0151 $\pm$ 0.0007 & 0.398 $\pm$ 0.020 & $-$3.27 $\pm$ 0.21 & 16.6 $\pm$ 0.7 \\
SIG & 30 & $-$0.0106 $\pm$ 0.0027 & 0.245 $\pm$ 0.080 & $-$1.54 $\pm$ 0.81 & 10.0 $\pm$ 2.7 \\
SIP & 30 & 0.0110 $\pm$ 0.0061 & $-$0.401 $\pm$ 0.184 & 4.85 $\pm$ 1.84 & $-$11.0 $\pm$ 6.1 \\
SIT & 30 & $-$0.3614 $\pm$ 0.1485 & 11.220 $\pm$ 4.606 & $-$115.90 $\pm$ 47.64 & 406.9 $\pm$ 164.2\\
\hline
\end{tabular}
\tablefoot{Parameters of the third order fit to the running median for all SF galaxies in the DP17, Centrals-CS, SIG samples, and central galaxies in the SIP and SIT samples, used to obtain their corresponding  12+log(O/H)--M$_\star$ relation, as shown in the right panel of Fig.~\ref{Fig:sfr_m}. The uncertainties of the parameters have been estimated from the square root of the diagonal of the covariance matrix of the parameter estimates using the \texttt{curve\_fit} function from the SciPy python library.}
\end{table*}

\subsection{Fundamental relation}
\label{subsec:fundamental_results}

\begin{figure*}
\centering
\includegraphics[width=.49\textwidth]{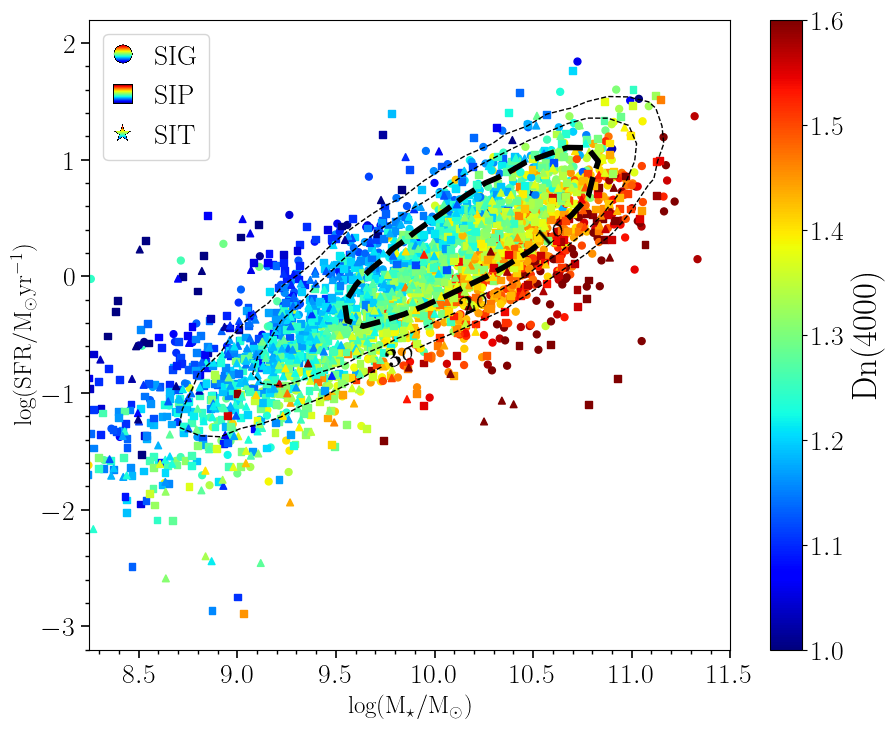}
\includegraphics[width=.49\textwidth]{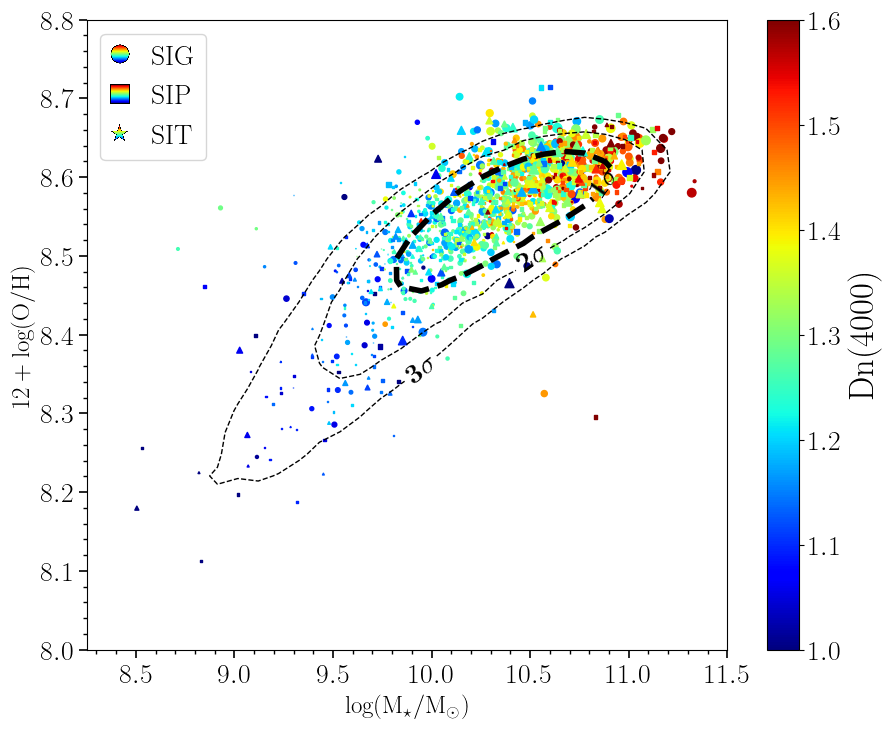}
\caption{Fundamental relation (SFR-M$_\star$-O/H) for central star forming galaxies with $8.5\,\leq\,\rm log\,(M_{\star}/M_{\odot})\,\leq\,11.5$ colour coded by the galaxy $D_n$(4000). The marker size in the left panel is proportional to the 12+log(O/H) values, while the marker size in the right panel is proportional to the SFR values. For reference, contours show the location of star-forming galaxies in the DP17 sample.}
\label{Fig:fundamental}
\end{figure*}

After deriving the SFR corrected for aperture effects and the oxygen abundance for our whole sample, we show the stellar mass -- metallicity -- SFR relation (MZSFR) in the panels of Fig.~\ref{Fig:fundamental} for star-forming galaxies in the SIG (circles), and central galaxies in the SIP (squares) and SIT (triangles), with $8.5\,\leq\,\rm log\,(M_{\star}/M_{\odot})\,\leq\,11.5$. From this figure, it can be seen that galaxies are located in the expected zone in this diagram. There are no outlier galaxies in this relation for SIG, SIP, and SIT galaxies. In addition, we found that the MZSFR is well represented by the first-order fit to the three-dimensional surface defined by the values of the M$_\star$, 12+log(O/H), and SFR in the form of $\rm 12+\log(O/H)~=~a\,\log(M_\star)~+~b\,\log(SFR)~+~ c$. The parameters of the fits, with their corresponding uncertainties, are shown in Table~\ref{tab:unZMRSFR}.

\begin{table*}[]
\centering
\caption{\label{tab:unZMRSFR}Parameters of the MZRSFR for SF galaxies in this study.}
\begin{tabular}{lccc}
\hline
\hline \\[-2ex]
 (1) & (2) & (3) & (4)  \\
sample & a & b & c \\
 & [$\rm \log(M_{\odot})^{-1}$] & [$\rm \log(M_{\odot}\,yr^{-1})^{-1}$] & \\
\hline
DP17 & 0.19 $\pm$ 0.02 & $-$0.02 $\pm$ 0.02 & 6.5 $\pm$ 0.2 \\
Centrals-CS & 0.14 $\pm$ 0.01 & 0.009 $\pm$ 0.009 & 7.1 $\pm$ 0.1 \\
SIG & 0.17 $\pm$ 0.02 & $-$0.00001 $\pm$ 0.00994  & 6.8 $\pm$ 0.2 \\
SIP & 0.17 $\pm$ 0.02 & $-$0.0005 $\pm$ 0.0103 & 6.8 $\pm$  0.2 \\
SIT & 0.11 $\pm$ 0.01 & 0.04 $\pm$ 0.01 & 7.3 $\pm$ 0.1 \\
\hline
\end{tabular}
\tablefoot{Parameters of the first order fit to the surface defined by the M$_\star$--12+log(O/H)--SFR values for all SF galaxies in the in the DP17, Centrals-CS, and SIG samples, and central SF galaxies in the SIP and SIT samples. The uncertainties of the parameters have been estimated using MCMC method with 100 mock models within the mean errors of the stellar mass, the SFR, and the oxygen abundances of each sample.}
\end{table*}

\subsection{Effect of the local environment and galactic conformity}
\label{sec:local_result}

\begin{figure*}
\includegraphics[width=\textwidth]{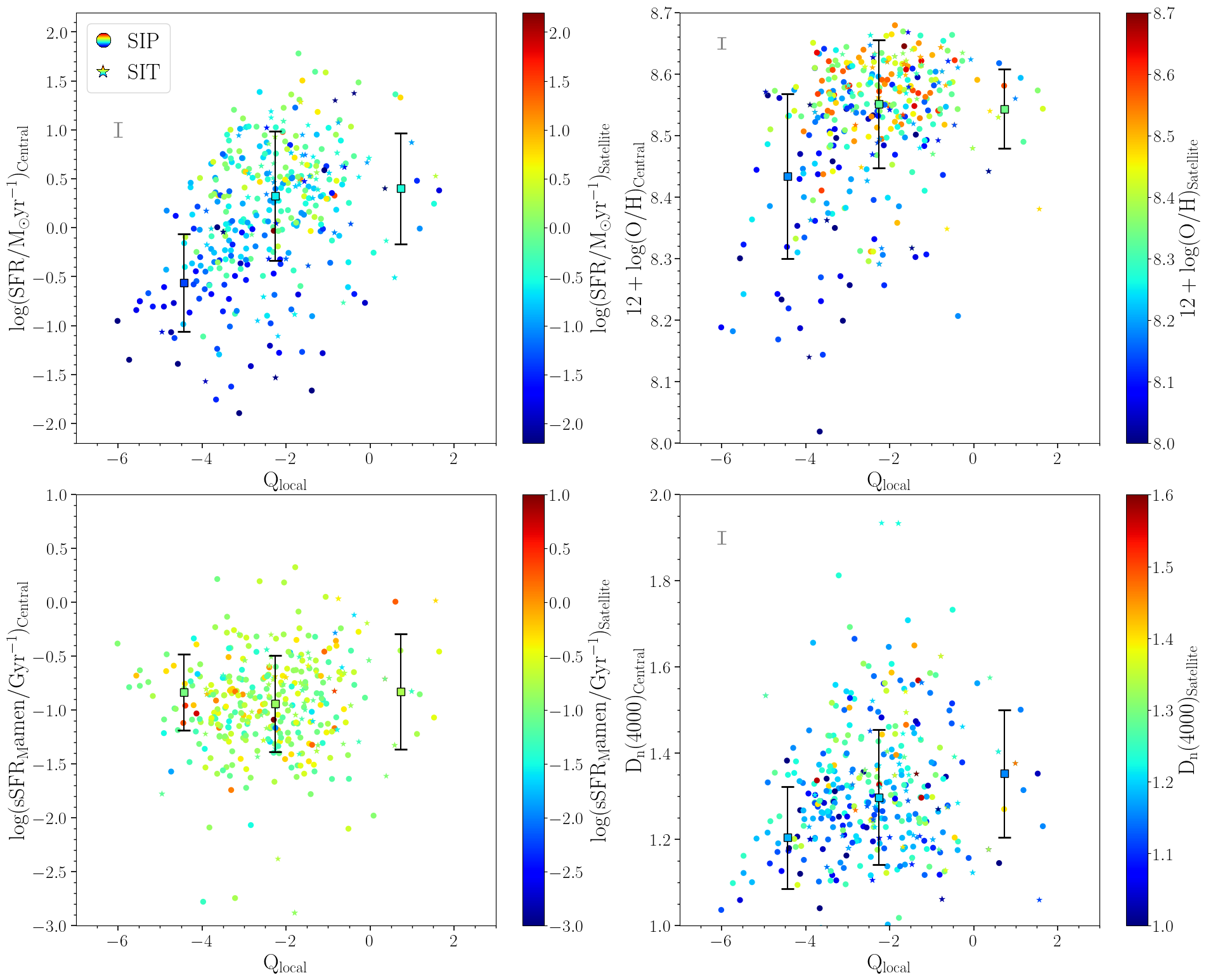}
\caption{Effect of the environment on the SFR, 12+log(O/H), sSFR, and $D_n$(4000) for star forming galaxies. Local tidal strength affecting the central galaxies in the isolated pairs (circles) and triplets (stars) versus: 1) the SFR of the central galaxy, coloured by the SFR of the satellite galaxy (left upper panel); 2) the 12+log(O/H) of the central galaxy, coloured by the 12+log(O/H) of the satellite galaxy (right upper panel); 3) the sSFR of the central galaxy, coloured by the sSFR of the satellite galaxy (left lower panel); and 4) the $D_n$(4000) of the central galaxy, coloured by the $D_n$(4000) of the satellite galaxy (right upper panel). Squares and error bars represent the median trend (coloured according to the colour bar of each panel) and dispersion, respectively, in three bins of the local tidal strength parameter (Q$\rm_{local}~\leq~-$4, $-$4~<~Q$\rm_{local}\leq$~0, and Q$\rm_{local}$~>~0). In the upper left on each panel we present a representative error, given by the median value of the error of the SFR (multiplied by 10 for better visualisation), 12+log(O/H), sSFR, and $D_n$(4000) for central galaxies. The tidal strength parameter values from \citep{2015A&A...578A.110A} do not have an associated error.}
\label{Fig:sfrOH_qpair}
\end{figure*}

\begin{figure*}
\includegraphics[width=\textwidth]{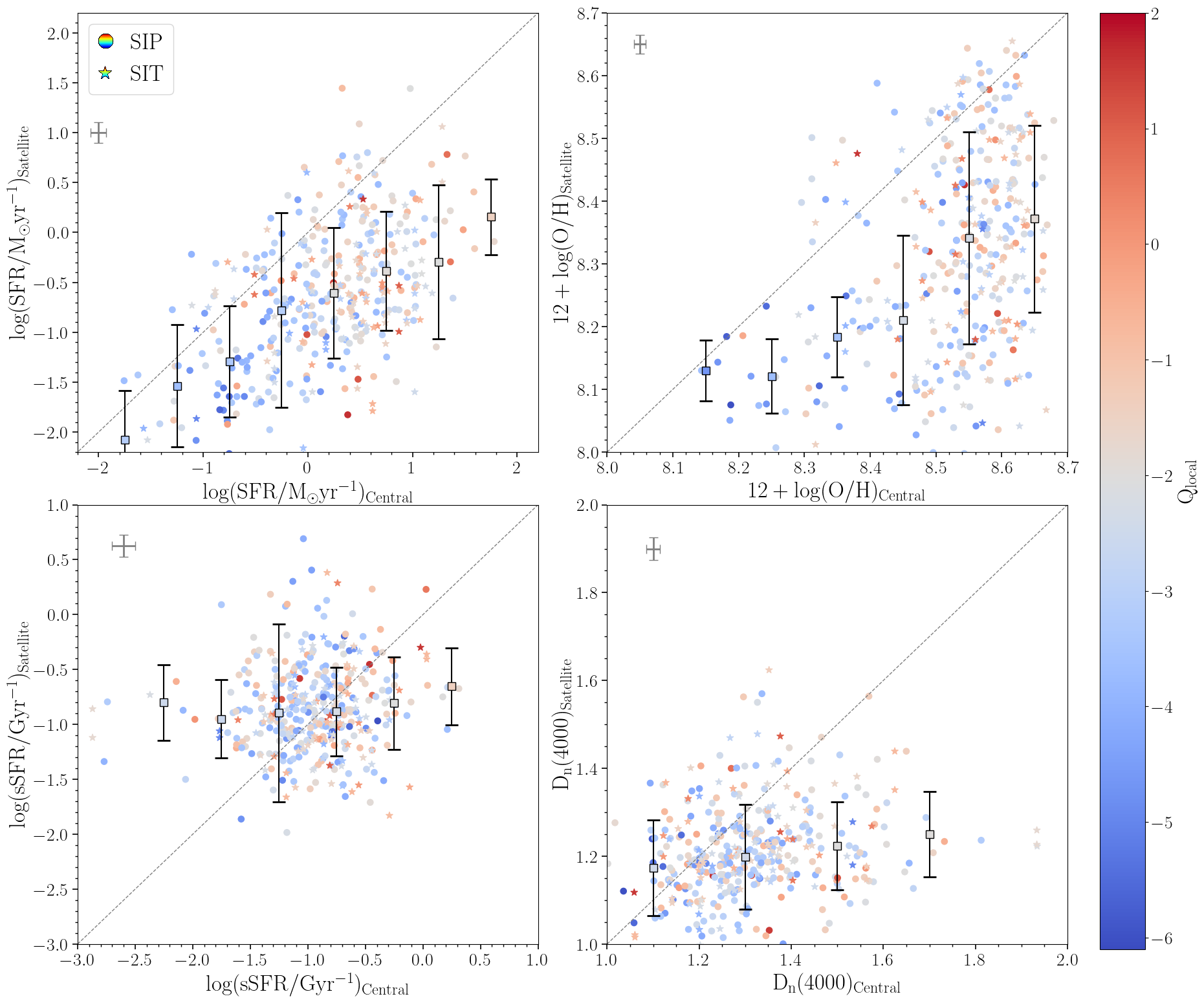}
\caption{Galactic conformity on the SFR, 12+log(O/H), sSFR, and $D_n$(4000) for star forming galaxies, coloured by the local tidal strength affecting the central galaxies in the isolated pairs (circles) and isolated triplets (stars). Upper left panel: the SFR of the central galaxy versus the SFR of the satellite galaxy. Upper right panel: 12+log(O/H) of the central galaxy versus 12+log(O/H) of the satellite galaxy. Lower left panel: sSFR of the central galaxy, versus the sSFR of the satellite galaxy. Lower right panel: $D_n$(4000) of the central galaxy versus $D_n$(4000) of the satellite galaxy. Grey dashed line is 1:1 relationship. Squares and error bars represent the median trend (coloured according to the colour bar) and dispersion, respectively, in three bins of the SFR, 12+log(O/H), sSFR, and $D_n$(4000). In the upper left on each panel we present a representative error, given by the median value of the error of the property in each panel for central and satellite galaxies. The median value of the uncertainties of the SFR is multiplied by 10 for better visualisation.}
\label{Fig:sfrOH_qpair_2}
\end{figure*}

To explore the effect of the local environment in the above relations we use the tidal strength parameter described in Sect.~\ref{subsec:quantification_environment} on central galaxies in the SIP and SIT. 
Fig.~\ref{Fig:sfrOH_qpair} shows the relation between the SFR (upper left panel), 12+log(O/H) (upper right panel), sSFR (lower left panel), and $D_n$(4000) (lower right panel) of the central galaxy in the SIP and SIT with the tidal strength parameter Q$\rm_{local}$, colour-coded by the value of the same property for the corresponding satellite galaxy. Complementarily, Fig.~\ref{Fig:sfrOH_qpair_2} shows the relation between these four properties for central and satellite galaxies, colour-coded by the Q$\rm_{local}$ parameter. 

To understand the differences between central and satellite galaxies, in the third row panels of Fig.~\ref{Fig:hists} we explore the distribution of the oxygen abundance considering all star-forming galaxies in the DP17, Centrals-CS, and SIG samples, in addition to: i) all star-forming galaxies in the SIP and SIT (left panel); ii) only the central galaxies in the SIP and SIT (central panel); and iii) only the satellite galaxies in the SIP and SIT (right panel). The median values for each galaxy sample are shown in Table~\ref{tab:meanerrprop} and are marked with different line styles and colours according to the legend.

\section{Discussion}
\label{sec:discu}

\subsection{SFR as a function of the M\texorpdfstring{$_\star$}{star}}
\label{sec:sfrmass_discussion}

From Fig.~\ref{Fig:sfr_m} we show that, at a fixed stellar mass, the SIG SF galaxies have lower SFR values ($\sim$0.2\,dex) than the MS of the sample of SDSS SF galaxies from \cite{2017A&A...599A..71D}. According to \citet{2022A&A...666A.186D} and \citep{2025arXiv250210078V}, the difference with the MS ($\Delta \rm MS$) is related to the age of the stellar population of the galaxies (measured by the $D_n$(4000) parameter), being galaxies above the MS younger than below the MS. Therefore SF SIG galaxies are systematically older than SF centrals and SDSS galaxies. Indeed, the lower panels in Fig.~\ref{Fig:hists} show that the median value of the $D_n$(4000) for SIG galaxies (1.34\,$\pm$\,0.02) is slightly higher than for DP17 galaxies (1.28\,$\pm$\,0.03), as the median values are nearly equivalent within the error margins. Therefore, SF isolated galaxies are older than SF SDSS galaxies on average. When we focus on the SIP and SIT galaxies the difference of the SFR at fixed stellar mass become smaller, with the values matching those of the MS, due to the contribution of satellite galaxies, which in general have lower SFR and $D_n$(4000) values than central galaxies. When considering central galaxies, the mean and median values of the $D_n$(4000) parameter are comparable, however, as shown in Fig.~\ref{Fig:sfr_m}, the effect of the addition of one companion galaxy is noticeable in the MS defined for SF galaxies in the SIG, SIP and SIT, with increasing SFR values at fixed stellar mass.

SIG galaxies have not interacted with another galaxy for at least 5\,Gyr \citep{2007A&A...470..505V}, therefore the secular processes of the galaxy dominate their evolution. These galaxies have not had stellar starburst produced by interactions with other galaxies and the gas is consumed more slowly, therefore their stellar population do not have undergone any strong acceleration processes in their recent lifetime. When we focus on SIP or SIT star-forming galaxies we can see that the SFR values are higher at a fixed mass (both in central and satellite galaxies). The successive interactions between the galaxies that form these pairs and triplets have enhanced the star formation. 

\begin{figure}
\includegraphics[width=\columnwidth]{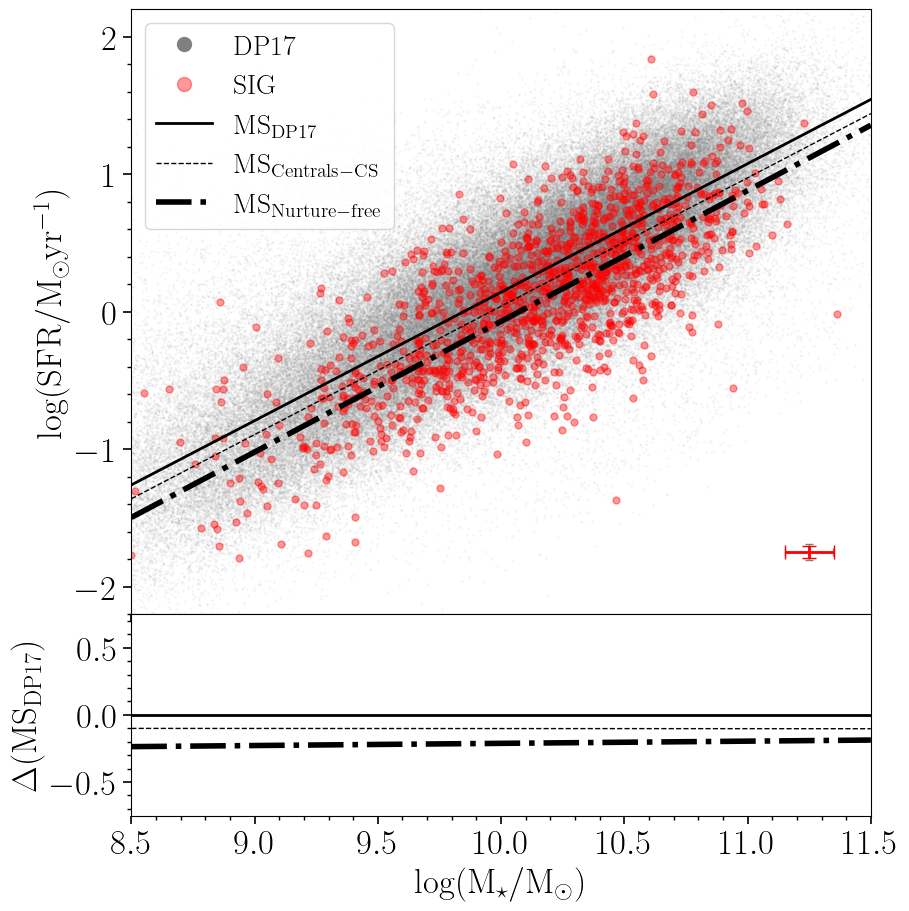}
\caption{Nurture-free main sequence defined for isolated star-forming galaxies with $8.5\,\leq\,\rm log\,(M_{\star}/M_{\odot})\,\leq\,11.5$. In addition, we represent the location of isolated star-forming galaxies in the diagram (red points). For comparison, we also show the relation for star forming galaxies in \citet{2017A&A...599A..71D} and the control sample (black continuos line and black dashed line, respectively), with grey points showing the values for star-forming galaxies in the DP17 sample. In the lower right corner we present a representative error, given by the median value of the error of the $\rm log\,SFR$ (multiplied by 10 for better visualisation) for each sample following the colours in the legends, and the median error of the $\rm log\,M_{\star}$ for galaxies in the comparison sample from the MPA catalogue, which is in agreement with the error we consider for galaxies in the SIG, SIP, and SIT samples ($\rm log\,(M_{\star}/M_{\odot})\,=\,0.10$\,dex). The lower panel show the average difference with the Main Sequence in \citet{2017A&A...599A..71D}.}
\label{Fig:nfMS}
\end{figure}

We also found that the SFR-M$_\star$ relation for SF isolated galaxies is tighter than SF SDSS galaxies. The scatter for isolated galaxies is 0.38\,dex meanwhile the scatter in the DP17 sample is 0.45\,dex. The values for SIP and SIT galaxies are also comparable with SDSS galaxies (0.41\,dex and 0.44\,dex, respectively). The scatter of the MS for SF galaxies is therefore affected by the local environment. The relation for the MS we obtained for isolated (SIG) SF galaxies in the local universe is:

\begin{equation}\label{MS-SIG}
   \rm \log(SFR)~=~a\,\log(M_\star) \, +\, b\quad,
\end{equation}
with a\,=\,0.952$\pm$0.005 and b\,=\,$-$9.586$\pm$0.052, as shown in Table~\ref{tab:unMS}.
This relation, presented in Fig.~\ref{Fig:nfMS}, slightly differs from the MS defined by \cite{2017A&A...599A..71D}, where the small differences account for the effects of the local environment. We therefore propose this relation as a ground level `nurture' free MS for SF galaxies in the local universe.

\subsection{Oxygen abundance as a function of the M\texorpdfstring{$_\star$}{star}}
\label{subsec:ohmass_discussion}

SIG star-forming galaxies are located at the top of the MZR relation and can be considered as metal-rich galaxies. From Fig.~\ref{Fig:hists} we show that the SIG galaxies present higher values, on average, than the sample of star-forming galaxies from \cite{2017A&A...599A..71D}, the SIP and SIT galaxies (considering both central and satellite galaxies) with mean values of 8.53\,$\pm$\,0.01\,dex, 8.46\,$\pm$\,0.02\,dex, 8.41\,$\pm$\,0.02\,dex and 8.38\,$\pm$\,0.02\,dex, respectively (as shown in Table~\ref{tab:meanerrprop}). The difference in the oxygen abundance derived between the SIG galaxies and the SDSS galaxies is very small ($\sim$0.05\,dex) and is within the error range of the calibrators used to derive the oxygen abundance. This result is similar to that found in previous studies \citep[e.g.][]{2017MNRAS.465.1358P}. When we separate the central and satellite galaxies in the SIP and SIT systems, as presented in the central and right panels of Fig.~\ref{Fig:hists}, we find that the central galaxies, in SIP and SIT, show similar oxygen abundance values to those in the SIG galaxies and galaxies in the Central-CS, and therefore similar to the values obtained for field galaxies in the \cite{2017A&A...599A..71D} sample. However, the satellite galaxies show lower oxygen abundance values ($\sim$0.18\,dex) than central SF galaxies. These differences are larger than the proper uncertainty in the calibrators, and also larger than the difference found by \cite{2008ApJ...672L.107E} (0.05–0.1\,dex). Therefore the difference we see in the left panel between the SIG galaxies with galaxies in the SIP and SIT is due to the properties of the satellite galaxies. 

In summary, central galaxies present slightly higher abundances than SDSS galaxies, and satellite galaxies have significantly lower abundances. However, the differences that we observe in the MZR for SIP and SIT galaxies with respect to SDSS galaxies are mainly due to the stellar mass, being satellite galaxies in general less massive than central galaxies. The dispersion with respect to the MZR is smaller in isolated SF galaxies (0.06\,dex) with respect to SDSS SF galaxies (0.09\,dex), pairs, and triplets (0.08\,dex for both samples). When separating SIP and SIT in central and satellite galaxies, the dispersion in the MZR for centrals is similar to SIG galaxies (0.06\,dex on average), while the dispersion for satellites is closer to SDSS galaxies (0.08\,dex on average). Even if the MZR for SIG galaxies is also tighter than for SDSS SF galaxies, as in the case of the SFR-M$_\star$ relation, the differences are too small to be related to the local environment, and more sensitive to stellar mass. Therefore, SF galaxies in isolated systems follows the MZR for SF galaxies in the local universe. Moreover, we do not have galaxies in the `outlier' subsample of galaxies found by \citet{2022A&A...666A.186D} that populate the youngest part of the 12+log(O/H)--M$_\star$ diagram in all the stellar mass range, from dwarf galaxies to Milky-Way like stellar mass galaxies.

\subsection{Fundamental relation}
\label{subsec:fundamental_discussion}

After visiting the stellar mass -- SFR relation and the stellar mass -- metallicity relation for star-forming galaxies in the SIG, SIP, and SIT, we explored the stellar mass -- metallicity -- SFR relation (MZSFR), known as the fundamental relation for star-forming galaxies, as presented in Fig.~\ref{Fig:fundamental}. We do not find that galaxies in isolated systems are located in a particular area, as we do not find either outlier galaxies. As expected, more massive galaxies in the three samples show higher values of SFR and 12+log(O/H). Less massive galaxies, which in general are satellite galaxies in the SIP and SIT, show lower values of the SFR and oxygen abundance, populating that corresponding region of the MZSFR. 

When exploring the three-dimensional MZSFR diagram as a function of the $D_n$(4000), as shown in Fig.~\ref{Fig:fundamental}, we confirm the results of \citet{2022A&A...666A.186D} for star-forming galaxies in the SDSS, where the MZSFR is found to be modelled by the age of the stellar populations. As expected, we find that galaxies with younger stellar populations have high SFR values at fixed 12+log(O/H) while galaxies with older stellar populations have lower SFR values at fixed 12+log(O/H). This correlation is also expected since $D_n$(4000) is also a parameter sensitive to the recent star formation history \citep{2023MNRAS.519.1149B}.

As originally found by \citet{2010MNRAS.408.2115M, 2010A&A...521L..53L}, and confirmed by simulations \citep{2012MNRAS.422..215Y,2016MNRAS.459.2632L}, the fundamental relation between the stellar mass, the SFR, and the oxygen abundances defines a three-dimensional surface. The first-order fit to the data to a plane in the form of 12+log(O/H)~=~a\,log(M$_\star$)~+~b\,log(SFR)~+~c (with a, b, and c constants) shows that the plane defined by isolated galaxies and central galaxies in pairs is similar (see Tab.~\ref{tab:unZMRSFR}), being the planes less steep than the planes defined by central SIT galaxies, and galaxies in the Centrals-CS and DP17 sample. the similarity between the plane defined by the SIG and SIP samples is explained by their similar MZR relations. However, we also find that the scatter of the MZSFR relation for SIG galaxies is the lowest with respect to any other sample, which might be related to the fact of showing also a tighter scatter in its MS relation.
Considering that central galaxies have a similar stellar mass distribution (as shown in the upper panels of Fig.~\ref{Fig:hists}), we interpret that the difference in the slope between the plane defined by SDSS galaxies is due to the influences of their local and large-scale environments. Hence, as we did for the SFR-M$_\star$ relation for isolated SIG star-forming galaxies, we propose a `nurture' free fundamental plane for galaxies evolving evolving secularly, in an equilibrium between gas inflow, outflow and star formation. This relation is defined as:

\begin{equation}\label{MS-SIG}
  \rm  12+\log(O/H)~=~a\,\log(M_\star)~+~b\,\log(SFR)~+~ c\quad,
\end{equation}
with a\,=\,0.17\,$\pm$\,0.02, b\,=\,$-$0.00001\,$\pm$\,0.00994, and c\,=\,6.8\,$\pm$\,0.2, as shown in Table~\ref{tab:unZMRSFR}. The uncertainties of the coefficients has been calculated using MCMC method within the mean errors of the stellar mass (0.1 dex), the SFR (0.008 dex), and the oxygen abundances (0.02 dex).

When performing the second-order fit to the data for SDSS galaxies, this shows curve shapes towards galaxies with high SFR and low 12+log(O/H) values in the low-mass and high-mass ends. These shapes are not found in the fits for SF galaxies in the SIG, SIP, and SIT samples, confirming our previous results: galaxies in isolated systems do not populate the subsample of `outlier' galaxies found by \citet{2022A&A...666A.186D}.

\subsection{Effect of the local environment and galactic conformity}
\label{subsec:local_discussion}

To explore the effect of the environment and galactic conformity in terms of the properties studied in this work, and to improve the statistical significance, we presented the results as a function of central versus satellite considering the SIP and SIT samples together. In general, there is not a strong correlation between the SFR (or the sSFR), the oxygen abundance, and the $D_n$(4000) with the local environment, as shown in Fig.~\ref{Fig:sfrOH_qpair}. However, central galaxies with the lowest values of Q$\rm_{local}$ (<$-$4) also present low values of the SFR (mean log(SFR)~=~$-$0.57) and 12+log(O/H) (mean value $\sim$8.41), in comparison with centrals with Q$\rm_{local}$~>~0 (mean log(SFR)~=~0.43 and mean 12+log(O/H)~=~8.57, respectively). In case of satellite galaxies, the mean SFR increases from log(SFR)~=~$-$1.05 to log(SFR)~=~$-$0.46, and the mean oxygen abundance from 12+log(O/H)~=~8.23 to 12+log(O/H)~=~8.45, when Q$\rm_{local}$~<~$-$4 and Q$\rm_{local}$~>~0, respectively. In a companion paper (Duarte Puertas et al., in prep.) we have found that galaxies with lower values of Q$\rm_{local}$  have higher values of projected distance between central and satellite galaxies and, therefore, are more isolated. On the other hand, according to \citet{2023A&A...670A..63V}, galaxies in isolated systems with Q$\rm_{local}$\,>\,-2 shows signs of interactions. These systems are therefore mainly composed of central galaxies with high values of the SFR and gas-phase metallicity (8.5\,<\,12+log(O/H)\,<\,8.6).

In terms of galactic conformity, there is a correlation between the SFR of the central galaxy and its satellite. Low SFR centrals have low SFR satellites (in particular systems with low Q$\rm_{local}$), and high SFR centrals have high SFR satellites, where the SFR of the satellites is in general lower than central galaxies in all cases. In other words, in the upper left panel of Fig.~\ref{Fig:sfrOH_qpair_2} we show that, for most isolated systems where both members are star-forming galaxies, the SFR of the central galaxy is higher than the one of the satellite galaxy. This correlation disappears when normalising by the stellar mass of the galaxies (sSFR). Therefore, in terms of SFR, central and satellites are disturbed in the same proportion with respect to their own mass.

We observe two trends when considering oxygen abundances of the central galaxy. Systems with low Q$\rm_{local}$ are in general composed of centrals with low 12+log(O/H) (<8.4 dex) and slightly lower satellites (<8.3 dex). On the other hand, centrals with high 12+log(O/H) values (> 8.4 dex) have all ranges of values for satellites, independently on the  tidal strength parameter of the system. Looking at the oxygen abundances of the satellites, satellite galaxies with 12+log(O/H)\,<\,8.3 dex have central galaxies presenting values in the range of 8.1 dex to solar\footnote{Solar metallicity: 12 + log(O/H) = 8.69 \citep{2009ARA&A..47..481A}.}, while for satellite galaxies with 12+log(O/H)\,>\,8.3 dex, the central galaxies show a nearly constant value around the solar value. These trends are observable in both, pairs (289 central-satellite associations) and triplets (106 central-satellite associations), therefore to show these with a larger statistic, we consider the two isolated systems for a total of 395 centrals and their respective satellite. 

\subsubsection{Galactic conformity for all BPT galaxies}

\begin{figure*}
\includegraphics[width=\textwidth]{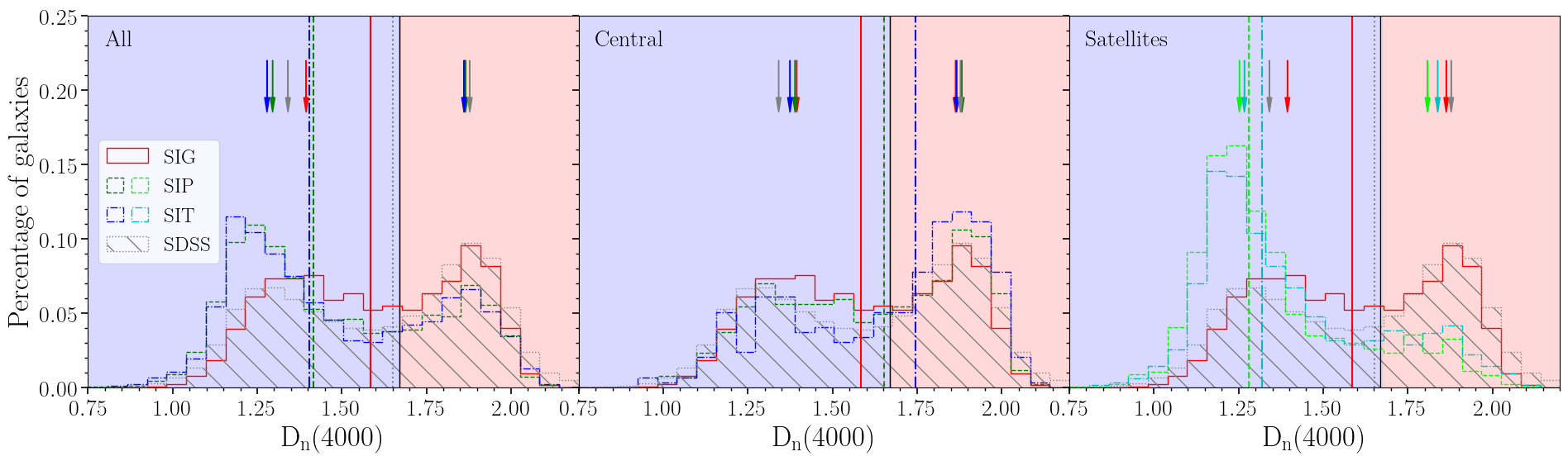}
\caption{Distributions of the $D_n$(4000) for all galaxies (star-forming, composite, and AGN), following the format in  Fig.~\ref{Fig:hists}. Vertical lines indicate the median values for all the galaxies in each sample, with the respective colour and line style according to the legend. Arrows indicate the median values when separating into young and old stellar populations (blue and red areas, respectively). The corresponding median values and standard deviations are presented in Table~\ref{tab:D4000}.}
\label{Fig:D4000}
\end{figure*}

In concordance with our previous findings, we do not observe either any correlation for the $D_n$(4000) parameter. This might be due to the fact that we are focusing on star-forming galaxies only, and then central and satellites have in general young stellar populations. Therefore, to explore the galactic conformity in isolated systems in terms of the age of the stellar populations, we consider the $D_n$(4000) for all central and their respective satellites, independently on their classification in the BPT diagram. The distribution for galaxies in each sample is shown in Fig.~\ref{Fig:D4000}, where Table~\ref{tab:D4000} shows the median values when separating into young and old stellar populations. As is also presented in the table, the fraction of young galaxies is in general higher than the fraction of old galaxies in all the samples and subsamples (central and satellite galaxies), except in central SIT galaxies, where the fraction of old galaxies is about 15\% higher than young galaxies. The difference between the fraction of young and old galaxies is minimal in the comparison sample ($\sim\,3\%$ only), where the difference in the SIG and central galaxies in the SIP is about 15\% and 3.5\%, respectively. This indicates that when considering all the galaxies (independently of their position in the BPT diagram), there are more young galaxies in the SIG than in the comparison sample, however star-forming isolated galaxies are systematically older, considering our previous results regarding the stellar mass -- SFR relation.

\begin{figure*}
\includegraphics[width=\textwidth]{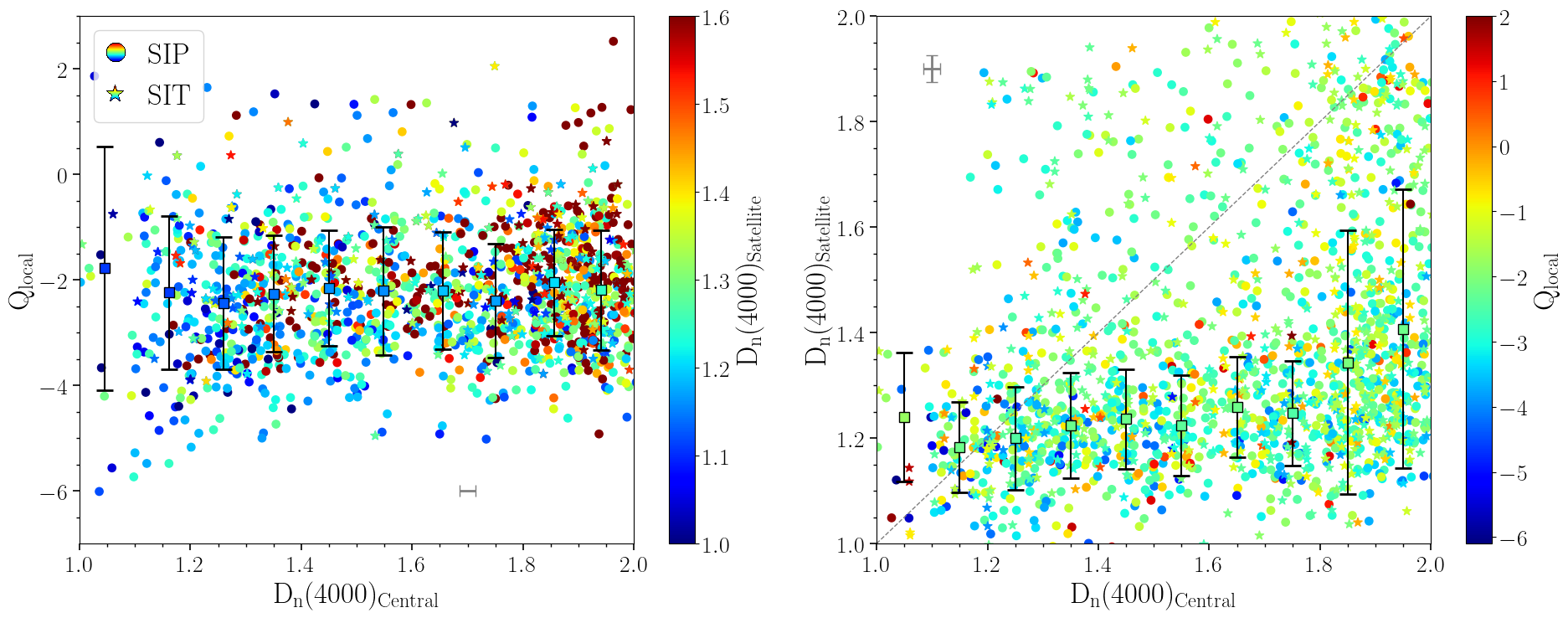}
\caption{Left panel: Effect of the environment on the $D_n$(4000) as in Fig.~\ref{Fig:sfrOH_qpair} but for for all the galaxies in the samples (star-forming, composite, and AGN). Right panel: Galactic conformity on the $D_n$(4000) as in Fig.~\ref{Fig:sfrOH_qpair_2} but for for all the galaxies in the samples (star-forming, composite, and AGN). The median trends and dispersion are shown as a function of the $D_n$(4000) of the central galaxy (squares and error bars, respectively coloured according to the corresponding colour bar).}
\label{Fig:D4000-censat}
\end{figure*}

In Fig.~\ref{Fig:D4000-censat} we show the $D_n$(4000) of central galaxies with the local environment, coloured by the $D_n$(4000) of satellite galaxies (left panel), and the $D_n$(4000) of central galaxies with the $D_n$(4000) of its corresponding satellite, coloured by the local environment (right panel). We do not find any relation with the Q$\rm_{local}$ parameter. Regarding galactic conformity, we find that the $D_n$(4000) of satellites is in general lower than central, with values typically under 1.4\,dex, and a wide range of values for central galaxies. Moreover, we find that in general the difference in the fraction between young and old galaxies is larger in satellite than in central galaxies (as shown in Table~\ref{tab:D4000}). The fraction of young satellites in the SIP is 72\% larger than the fraction of old satellites, and about 60\% in the SIT. These trends are expected by sample selection, where the central galaxy is generally more massive than satellites, and therefore older and presenting more early-type morphologies (Duarte-Puertas et al., in preparation). Systems with low values of the $D_n$(4000) for both, central and satellite, are found in systems with intermediate ($\sim$\,$-$2) and extreme (<\,$-$5 and >\,0) Q$\rm_{local}$ values. The median $D_n$(4000) for SIP and SIT centrals is higher than SIG galaxies (0.06 and 0.16 dex, respectively, as shown in the central panel of Fig.~\ref{Fig:D4000}), therefore isolated galaxies are slightly younger than central galaxies in isolated pairs and isolated triplets. However when separating into young and old stellar populations (see Table~\ref{tab:D4000}), the difference between centrals and isolated galaxies are much smaller, with young centrals slightly younger than young isolated galaxies, even if the fraction of young central galaxies in the SIT is is lower. Systems with Q$\rm_{local}$\,>\,$-$2 and low values of $D_n$(4000) for central and satellite, might be then the case where interaction with the satellite is rejuvenating the stellar populations of the galaxies, as suggested in \citet{2023A&A...670A..63V} for isolated triplets. In the case of systems with Q$\rm_{local}$\,<\,$-$4 (systems with largest protected separations between central and satellites), these are mainly isolated pairs with satellites with mean $D_n$(4000)~=~1.2, for which a mean stellar age younger than 150\,Myr is expected \citep{2006MNRAS.370..721M}. It might be worthy to explore if these 'loose' isolated systems are also mainly found in isolated large-scale environments, as cosmic voids, for instance, and therefore their dynamic evolution is slower than in denser environments.

\begin{table*}[]
\centering
\caption{\label{tab:D4000}Median values of the distributions of D$_n$(4000) parameter for all the galaxies in this study.}
\begin{tabular}{lcccccc}
\hline
\hline \\[-2ex]
 (1) & (2) & (3) & (4) & (5) & (6) & (7)\\
sample & subsample & all & young & old & $\rm f_{young}$ & $\rm f_{old}$\\
 & & & & & [\%] & [\%]\\ 
\hline
SDSS & -- & 1.65 $\pm$ 0.32 & 1.34 $\pm$ 0.21 & 1.88 $\pm$ 0.13 & 51.33 & 48.67 \\
\hline
SIG & -- & 1.59 $\pm$ 0.27 & 1.40 $\pm$ 0.16 & 1.86 $\pm$ 0.10 & 57.77 &  42.23 \\
\hline
SIP & -- & 1.42 $\pm$ 0.30 & 1.29 $\pm$ 0.18 & 1.87 $\pm$ 0.10 & 69.08 &  30.92 \\
 & central & 1.65 $\pm$ 0.28 & 1.36 $\pm$ 0.18 & 1.88 $\pm$ 0.10 & 51.77 & 48.23 \\
 & satellites &  1.28 $\pm$ 0.25 & 1.25 $\pm$ 0.16 & 1.81 $\pm$ 0.09 & 85.83 & 14.17 \\
\hline
SIT & -- & 1.41 $\pm$ 0.30 & 1.28 $\pm$ 0.16 & 1.86 $\pm$ 0.12 & 67.75 &  32.25 \\
 & central & 1.75 $\pm$ 0.28 & 1.37 $\pm$ 0.17 & 1.87 $\pm$ 0.10 & 42.37 & 57.63 \\
 & satellites & 1.32 $\pm$ 0.27 & 1.28 $\pm$ 0.16 & 1.86 $\pm$ 0.12 & 79.71 & 20.29 \\
\hline
\end{tabular}
\tablefoot{Median and standard deviation of the distribution of the $D_n$(4000) parameter for all the galaxies in the samples (star-forming, composite, and AGN), and divided into young ($D_n$(4000)\,$\leq$\,1.67) and old ($D_n$(4000)\,$>$\,1.67) galaxies. The last two columns correspond to the fraction of young and old galaxies in each sample, respectively, thus for the same sample $\rm f_{young}$ + $\rm f_{old}$ = 1.}
\end{table*}

\section{Summary and conclusions}
\label{sec:conclu}

In this study we establish the relations between stellar mass, gas-phase metallicity, and star-formation rate (i.e. the mass-metallicity, MZR, and the star-formation rate-mass relation, SFRM) for galaxies in the SDSS catalogue of Isolated Galaxies (SIG), SDSS catalogue of Isolated Pairs (SIP), and SDSS catalogue of Isolated Triplets (SIT) compiled by \citet{2015A&A...578A.110A}, to explore the influence of the environment on these relations in isolated systems in the local universe (with 0.005\,$\leq$\,z\,$\leq$\,0.080). 

The systems are isolated without neighbours in a volume of 1\,Mpc projected field radius and a line-of-sight velocity difference $\Delta v$\,=\,500\,km\,s$^{-1}$. \citet{2015A&A...578A.110A} explored the distribution of the projected distances (d) and $\Delta v$ of the galaxies in pairs and triplets to define physically bound isolated systems when $\Delta v$\,$\leq$\,160\,km\,s$^{-1}$ at projected distance d\,$\leq$\,450\,kpc. We use the tidal strength parameter, Q$\rm_{local}$, provided by \citet{2015A&A...578A.110A} for the galaxies in these samples to explore the effects of the local environment. 

We made use of public information of stellar masses and  emission line fluxes provided by Max-Planck-Institut für Astrophysik and Johns Hopkins University \citep[MPA-JHU,][]{2003MNRAS.341...33K,2004MNRAS.351.1151B,2004ApJ...613..898T,2007ApJS..173..267S} to select 1,282 star-forming isolated galaxies (out of 3,702 galaxies in the SIG), 1,209 star-forming galaxies isolated pairs (out of 2,480 galaxies in the SIP), and 460 star-forming galaxies in isolated triplets (out of 945 galaxies in the SIT).

We used aperture-corrected methods to derive the Star Formation Rate (SFR) in these galaxies according to \citet{2022A&A...666A.186D}. We derive the oxygen abundance using the empirical calibration S of \cite{2016MNRAS.457.3678P} from the [SII] emission line since the emission line $\rm [OII]{\lambda, 3727, 3729}$ is not covered in the optical SDSS spectra for galaxies with z~$<$~0.02. We also used the $D_n$(4000) spectral index, as defined in \citet{1999ApJ...527...54B}, as the stellar population age indicator \citep{2006MNRAS.370..721M}, which is small ($D_n$(4000)~$\leq$~1.67) for galaxies with younger stellar populations, and large ($D_n$(4000)~>~1.67) for older stellar populations. 

Our main conclusions are the following:

\begin{enumerate}
\item The SFRM for star-forming isolated galaxies shows a tighter correlation than the comparison sample of star-forming galaxies from the SDSS. We also find that, at fixed stellar mass, the SIG star-forming galaxies present lower SFR. This indicates that isolated galaxies have not had stellar starburst produced by interactions with other galaxies and the gas is consumed more slowly. Based on our results, we propose a ground level `nurture' free SFR-M$_\star$ relation for SF galaxies in the local universe.
\item The slope in the main sequence is comparable in all samples. However, the effect of the addition of one companion galaxy is noticeable in the SIG, SIP and SIT samples, with higher SFR values at fixed stellar mass. We also find that the successive interactions between galaxies that form isolated pairs and isolated triplets, both for central and satellites, have enhanced their star formation. 
\item When exploring the MZR relation, we do not find a significant difference between isolated galaxies and field galaxies. However, when we explore central and satellite galaxies in isolated pairs and triplets, we find that central galaxies present oxygen abundances comparable with isolated galaxies, being more metal-rich than satellites. Therefore, the average abundance content on central galaxies is not altered by the presence of their companion.
\item Star-forming galaxies in isolated galaxies, isolated pairs, and isolated triplets follow the fundamental relation, we have not found outlier galaxies. When exploring this relation as a function of the age, we find that galaxies with younger stellar populations have high SFR values at fixed 12+log(O/H), while galaxies with older stellar populations have lower SFR values at fixed 12+log(O/H).
\item In general, we do not find a strong correlation of these relations with the local environment in isolated pairs and isolated triplets. However, systems composed of central galaxies with high values of the SFR and gas-phase metallicity (8.5\,<\,12 + log(O/H)\,<\,8.6) present high tidal strength parameter (Q$\rm_{local}$~>~-2), which indicates that interactions might be happening between their member galaxies. The low values of $D_n$(4000) for central and satellite in these cases, might be indicating that the interaction with the satellite is rejuvenating the stellar populations of central galaxies.
\item For most isolated systems where both members are star-forming galaxies, the SFR of the central galaxy is higher than the one of the satellite galaxy. This correlation disappears when normalising by the stellar mass of the galaxies. This means that, in terms of SFR, central and satellites are disturbed in the same proportion with respect to their own mass.
\item Low oxygen abundance satellites in isolated pair and isolated triplets are found around all ranges of values for central galaxies, however high oxygen abundance satellites are found mainly surrounding high metallicity central galaxies. 
\end{enumerate}

Overall star-forming galaxies in isolated systems (isolated galaxies, isolated pairs, and isolated triplets) follow the expected correlations between SFR, stellar mass, and gas-phase metallicity for star-forming galaxies in the local universe. These correlations are tighter for isolated galaxies, therefore we propose a `nurture' free main sequence and fundamental plane for galaxies evolving evolving secularly, in an equilibrium between gas inflow, outflow and star formation. We do not find outlier galaxies in the fundamental relation. Star formation in isolated galaxies and central galaxies in isolated pairs and isolated triplets is mainly regulated by smooth secular processes. A slight increment of the SFR and metallicity is observed for both, central and satellite galaxies in isolated pairs and isolated triplets where interactions might be occurring in the systems.

\begin{acknowledgements}

The authors thank the referee for his/her constructive comments.
We acknowledge financial support from projects PID2020-113689GB-I00 and PID2023-149578NB-I00, financed by MCIN/AEI/10.13039/501100011033 and FEDER/UE; from FEDER/Junta de Andalucía-Consejería de Transformación Económica, Industria, Conocimiento y Universidades/Proyecto A-FQM-510-UGR20; from grant AST22\_4.4 financed by Junta de Andalucía-Consejería de Universidad, Investigación e Innovación and Gobierno de España and European Union-NextGenerationEU; and from projects P20\_00334 and FQM 108 financed by the Junta de Andalucía. 
M.A-F. acknowledges support from ANID FONDECYT iniciaci\'on project 11200107 and the Emergia program (EMERGIA20\_38888) from Consejer\'ia de Universidad, Investigaci\'on e Innovaci\'on de la Junta de Andaluc\'ia. 
S.D.P. acknowledges financial support from Juan de la Cierva Formaci\'on fellowship (FJC2021-047523-I) financed by MCIN/AEI/10.13039/501100011033 and by the European Union "NextGenerationEU"/PRTR, Ministerio de Econom\'ia y Competitividad under grant PID2019-107408GB-C44 and PID2022-136598NB-C32, and also from the Fonds de Recherche du Qu\'ebec - Nature et Technologies. \\

Funding for SDSS-III has been provided by the Alfred P. Sloan Foundation, the Participating Institutions, the National Science Foundation, and the U.S. Department of Energy Office of Science. The SDSS-III web site is http://www.sdss3.org/. SDSS-III is managed by the Astrophysical Research Consortium for the Participating Institutions of the SDSS-III Collaboration including the University of Arizona, the Brazilian Participation Group, Brookhaven National Laboratory, Carnegie Mellon University, University of Florida, the French Participation Group, the German Participation Group, Harvard University, the Instituto de Astrofisica de Canarias, the Michigan State/Notre Dame/JINA Participation Group, Johns Hopkins University, Lawrence Berkeley National Laboratory, Max Planck Institute for Astrophysics, Max Planck Institute for Extraterrestrial Physics, New Mexico State University, New York University, Ohio State University, Pennsylvania State University, University of Portsmouth, Princeton University, the Spanish Participation Group, University of Tokyo, University of Utah, Vanderbilt University, University of Virginia, University of Washington, and Yale University.\\

We acknowledge the use of STILTS and TOPCAT tools \citet{2005ASPC..347...29T}.\\

This research made use of Astropy, a community-developed core Python (http://www.python.org) package for Astronomy \citep{2013A&A...558A..33A, 2018AJ....156..123A}; ipython \citep{PER-GRA:2007}; matplotlib \citep{Hunter:2007}; SciPy, a collection of open source software for scientific computing in Python \citep{2020SciPy-NMeth}; pandas, an open source data analysis and manipulation tool \citep{reback2020pandas, mckinney-proc-scipy-2010}; NumPy, a structure for efficient numerical computation \citep{2011CSE....13b..22V}; and Uncertainties: a Python package for calculations with uncertainties, Eric O. Lebigot, \url{http://pythonhosted.org/uncertainties/}.

\end{acknowledgements}


\bibliography{paper}

\begin{thebibliography}{105}
\expandafter\ifx\csname natexlab\endcsname\relax\def\natexlab#1{#1}\fi

\bibitem[{{Ahn} {et~al.}(2014){Ahn}, {Alexandroff}, {Allende Prieto}, {Anders},
  {Anderson}, {Anderton}, {Andrews}, {Aubourg}, {Bailey}, {Bastien}, \&
  et~al.}]{2014ApJS..211...17A}
{Ahn}, C.~P., {Alexandroff}, R., {Allende Prieto}, C., {et~al.} 2014, \apjs,
  211, 17

\bibitem[{{Alam} {et~al.}(2015){Alam}, {Albareti}, {Allende Prieto}, {Anders},
  {Anderson}, {Anderton}, {Andrews}, {Armengaud}, {Aubourg}, {Bailey}, {Basu},
  {Bautista}, {Beaton}, {Beers}, {Bender}, {Berlind}, {Beutler}, {Bhardwaj},
  {Bird}, {Bizyaev}, {Blake}, {Blanton}, {Blomqvist}, {Bochanski}, {Bolton},
  {Bovy}, {Shelden Bradley}, {Brandt}, {Brauer}, {Brinkmann}, {Brown},
  {Brownstein}, {Burden}, {Burtin}, {Busca}, {Cai}, {Capozzi}, {Carnero
  Rosell}, {Carr}, {Carrera}, {Chambers}, {Chaplin}, {Chen}, {Chiappini},
  {Chojnowski}, {Chuang}, {Clerc}, {Comparat}, {Covey}, {Croft}, {Cuesta},
  {Cunha}, {da Costa}, {Da Rio}, {Davenport}, {Dawson}, {De Lee}, {Delubac},
  {Deshpande}, {Dhital}, {Dutra-Ferreira}, {Dwelly}, {Ealet}, {Ebelke},
  {Edmondson}, {Eisenstein}, {Ellsworth}, {Elsworth}, {Epstein}, {Eracleous},
  {Escoffier}, {Esposito}, {Evans}, {Fan}, {Fern{\'a}ndez-Alvar}, {Feuillet},
  {Filiz Ak}, {Finley}, {Finoguenov}, {Flaherty}, {Fleming}, {Font-Ribera},
  {Foster}, {Frinchaboy}, {Galbraith-Frew}, {Garc{\'\i}a},
  {Garc{\'\i}a-Hern{\'a}ndez}, {Garc{\'\i}a P{\'e}rez}, {Gaulme}, {Ge},
  {G{\'e}nova-Santos}, {Georgakakis}, {Ghezzi}, {Gillespie}, {Girardi},
  {Goddard}, {Gontcho}, {Gonz{\'a}lez Hern{\'a}ndez}, {Grebel}, {Green},
  {Grieb}, {Grieves}, {Gunn}, {Guo}, {Harding}, {Hasselquist}, {Hawley},
  {Hayden}, {Hearty}, {Hekker}, {Ho}, {Hogg}, {Holley-Bockelmann}, {Holtzman},
  {Honscheid}, {Huber}, {Huehnerhoff}, {Ivans}, {Jiang}, {Johnson},
  {Kinemuchi}, {Kirkby}, {Kitaura}, {Klaene}, {Knapp}, {Kneib}, {Koenig},
  {Lam}, {Lan}, {Lang}, {Laurent}, {Le Goff}, {Leauthaud}, {Lee}, {Lee},
  {Licquia}, {Liu}, {Long}, {L{\'o}pez-Corredoira}, {Lorenzo-Oliveira},
  {Lucatello}, {Lundgren}, {Lupton}, {Mack}, {Mahadevan}, {Maia}, {Majewski},
  {Malanushenko}, {Malanushenko}, {Manchado}, {Manera}, {Mao}, {Maraston},
  {Marchwinski}, {Margala}, {Martell}, {Martig}, {Masters}, {Mathur},
  {McBride}, {McGehee}, {McGreer}, {McMahon}, {M{\'e}nard}, {Menzel},
  {Merloni}, {M{\'e}sz{\'a}ros}, {Miller}, {Miralda-Escud{\'e}}, {Miyatake},
  {Montero-Dorta}, {More}, {Morganson}, {Morice-Atkinson}, {Morrison},
  {Mosser}, {Muna}, {Myers}, {Nandra}, {Newman}, {Neyrinck}, {Nguyen},
  {Nichol}, {Nidever}, {Noterdaeme}, {Nuza}, {O'Connell}, {O'Connell},
  {O'Connell}, {Ogando}, {Olmstead}, {Oravetz}, {Oravetz}, {Osumi}, {Owen},
  {Padgett}, {Padmanabhan}, {Paegert}, {Palanque-Delabrouille}, {Pan},
  {Parejko}, {P{\^a}ris}, {Park}, {Pattarakijwanich}, {Pellejero-Ibanez},
  {Pepper}, {Percival}, {P{\'e}rez-Fournon}, {P{\'e}rez-R{\`a}fols},
  {Petitjean}, {Pieri}, {Pinsonneault}, {Porto de Mello}, {Prada}, {Prakash},
  {Price-Whelan}, {Protopapas}, {Raddick}, {Rahman}, {Reid}, {Rich}, {Rix},
  {Robin}, {Rockosi}, {Rodrigues}, {Rodr{\'\i}guez-Torres}, {Roe}, {Ross},
  {Ross}, {Rossi}, {Ruan}, {Rubi{\~n}o-Mart{\'\i}n}, {Rykoff},
  {Salazar-Albornoz}, {Salvato}, {Samushia}, {S{\'a}nchez}, {Santiago},
  {Sayres}, {Schiavon}, {Schlegel}, {Schmidt}, {Schneider}, {Schultheis},
  {Schwope}, {Sc{\'o}ccola}, {Scott}, {Sellgren}, {Seo}, {Serenelli}, {Shane},
  {Shen}, {Shetrone}, {Shu}, {Silva Aguirre}, {Sivarani}, {Skrutskie},
  {Slosar}, {Smith}, {Sobreira}, {Souto}, {Stassun}, {Steinmetz}, {Stello},
  {Strauss}, {Streblyanska}, {Suzuki}, {Swanson}, {Tan}, {Tayar}, {Terrien},
  {Thakar}, {Thomas}, {Thomas}, {Thompson}, {Tinker}, {Tojeiro}, {Troup},
  {Vargas-Maga{\~n}a}, {Vazquez}, {Verde}, {Viel}, {Vogt}, {Wake}, {Wang},
  {Weaver}, {Weinberg}, {Weiner}, {White}, {Wilson}, {Wisniewski},
  {Wood-Vasey}, {Ye`che}, {York}, {Zakamska}, {Zamora}, {Zasowski}, {Zehavi},
  {Zhao}, {Zheng}, {Zhou}, {Zhou}, {Zou}, \& {Zhu}}]{2015ApJS..219...12A}
{Alam}, S., {Albareti}, F.~D., {Allende Prieto}, C., {et~al.} 2015, \apjs, 219,
  12

\bibitem[{{Andrews} \& {Martini}(2013)}]{2013ApJ...765..140A}
{Andrews}, B.~H. \& {Martini}, P. 2013, \apj, 765, 140

\bibitem[{{Argudo-Fern{\'a}ndez} {et~al.}(2015){Argudo-Fern{\'a}ndez},
  {Verley}, {Bergond}, {Duarte Puertas}, {Ramos Carmona}, {Sabater},
  {Fern{\'a}ndez Lorenzo}, {Espada}, {Sulentic}, {Ruiz}, \&
  {Leon}}]{2015A&A...578A.110A}
{Argudo-Fern{\'a}ndez}, M., {Verley}, S., {Bergond}, G., {et~al.} 2015, \aap,
  578, A110

\bibitem[{{Argudo-Fern{\'a}ndez} {et~al.}(2014){Argudo-Fern{\'a}ndez},
  {Verley}, {Bergond}, {Sulentic}, {Sabater}, {Fern{\'a}ndez Lorenzo},
  {Espada}, {Leon}, {S{\'a}nchez-Exp{\'o}sito}, {Santander-Vela}, \&
  {Verdes-Montenegro}}]{2014A&A...564A..94A}
{Argudo-Fern{\'a}ndez}, M., {Verley}, S., {Bergond}, G., {et~al.} 2014, \aap,
  564, A94

\bibitem[{{Argudo-Fern{\'a}ndez} {et~al.}(2013){Argudo-Fern{\'a}ndez},
  {Verley}, {Bergond}, {Sulentic}, {Sabater}, {Fern{\'a}ndez Lorenzo}, {Leon},
  {Espada}, {Verdes-Montenegro}, {Santander-Vela}, {Ruiz}, \&
  {S{\'a}nchez-Exp{\'o}sito}}]{2013A&A...560A...9A}
{Argudo-Fern{\'a}ndez}, M., {Verley}, S., {Bergond}, G., {et~al.} 2013, \aap,
  560, A9

\bibitem[{{Asplund} {et~al.}(2009){Asplund}, {Grevesse}, {Sauval}, \&
  {Scott}}]{2009ARA&A..47..481A}
{Asplund}, M., {Grevesse}, N., {Sauval}, A.~J., \& {Scott}, P. 2009, \araa, 47,
  481

\bibitem[{{Astropy Collaboration} {et~al.}(2018){Astropy Collaboration},
  {Price-Whelan}, {Sip{\H{o}}cz}, {G{\"u}nther}, {Lim}, {Crawford}, {Conseil},
  {Shupe}, {Craig}, {Dencheva}, {Ginsburg}, {Vand erPlas}, {Bradley},
  {P{\'e}rez-Su{\'a}rez}, {de Val-Borro}, {Aldcroft}, {Cruz}, {Robitaille},
  {Tollerud}, {Ardelean}, {Babej}, {Bach}, {Bachetti}, {Bakanov}, {Bamford},
  {Barentsen}, {Barmby}, {Baumbach}, {Berry}, {Biscani}, {Boquien}, {Bostroem},
  {Bouma}, {Brammer}, {Bray}, {Breytenbach}, {Buddelmeijer}, {Burke},
  {Calderone}, {Cano Rodr{\'\i}guez}, {Cara}, {Cardoso}, {Cheedella}, {Copin},
  {Corrales}, {Crichton}, {D'Avella}, {Deil}, {Depagne}, {Dietrich}, {Donath},
  {Droettboom}, {Earl}, {Erben}, {Fabbro}, {Ferreira}, {Finethy}, {Fox},
  {Garrison}, {Gibbons}, {Goldstein}, {Gommers}, {Greco}, {Greenfield},
  {Groener}, {Grollier}, {Hagen}, {Hirst}, {Homeier}, {Horton}, {Hosseinzadeh},
  {Hu}, {Hunkeler}, {Ivezi{\'c}}, {Jain}, {Jenness}, {Kanarek}, {Kendrew},
  {Kern}, {Kerzendorf}, {Khvalko}, {King}, {Kirkby}, {Kulkarni}, {Kumar},
  {Lee}, {Lenz}, {Littlefair}, {Ma}, {Macleod}, {Mastropietro}, {McCully},
  {Montagnac}, {Morris}, {Mueller}, {Mumford}, {Muna}, {Murphy}, {Nelson},
  {Nguyen}, {Ninan}, {N{\"o}the}, {Ogaz}, {Oh}, {Parejko}, {Parley}, {Pascual},
  {Patil}, {Patil}, {Plunkett}, {Prochaska}, {Rastogi}, {Reddy Janga},
  {Sabater}, {Sakurikar}, {Seifert}, {Sherbert}, {Sherwood-Taylor}, {Shih},
  {Sick}, {Silbiger}, {Singanamalla}, {Singer}, {Sladen}, {Sooley},
  {Sornarajah}, {Streicher}, {Teuben}, {Thomas}, {Tremblay}, {Turner},
  {Terr{\'o}n}, {van Kerkwijk}, {de la Vega}, {Watkins}, {Weaver}, {Whitmore},
  {Woillez}, {Zabalza}, \& {Astropy Contributors}}]{2018AJ....156..123A}
{Astropy Collaboration}, {Price-Whelan}, A.~M., {Sip{\H{o}}cz}, B.~M., {et~al.}
  2018, \aj, 156, 123

\bibitem[{{Astropy Collaboration} {et~al.}(2013){Astropy Collaboration},
  {Robitaille}, {Tollerud}, {Greenfield}, {Droettboom}, {Bray}, {Aldcroft},
  {Davis}, {Ginsburg}, {Price-Whelan}, {Kerzendorf}, {Conley}, {Crighton},
  {Barbary}, {Muna}, {Ferguson}, {Grollier}, {Parikh}, {Nair}, {Unther},
  {Deil}, {Woillez}, {Conseil}, {Kramer}, {Turner}, {Singer}, {Fox}, {Weaver},
  {Zabalza}, {Edwards}, {Azalee Bostroem}, {Burke}, {Casey}, {Crawford},
  {Dencheva}, {Ely}, {Jenness}, {Labrie}, {Lim}, {Pierfederici}, {Pontzen},
  {Ptak}, {Refsdal}, {Servillat}, \& {Streicher}}]{2013A&A...558A..33A}
{Astropy Collaboration}, {Robitaille}, T.~P., {Tollerud}, E.~J., {et~al.} 2013,
  \aap, 558, A33

\bibitem[{{Baker} {et~al.}(2023){Baker}, {Maiolino}, {Belfiore}, {Curti},
  {Bluck}, {Lin}, {Ellison}, {Thorp}, \& {Pan}}]{2023MNRAS.519.1149B}
{Baker}, W.~M., {Maiolino}, R., {Belfiore}, F., {et~al.} 2023, \mnras, 519,
  1149

\bibitem[{{Baldwin} {et~al.}(1981){Baldwin}, {Phillips}, \&
  {Terlevich}}]{1981PASP...93....5B}
{Baldwin}, J.~A., {Phillips}, M.~M., \& {Terlevich}, R. 1981, \pasp, 93, 5

\bibitem[{{Balogh} {et~al.}(1999){Balogh}, {Morris}, {Yee}, {Carlberg}, \&
  {Ellingson}}]{1999ApJ...527...54B}
{Balogh}, M.~L., {Morris}, S.~L., {Yee}, H.~K.~C., {Carlberg}, R.~G., \&
  {Ellingson}, E. 1999, \apj, 527, 54

\bibitem[{{Behroozi} {et~al.}(2010){Behroozi}, {Conroy}, \&
  {Wechsler}}]{2010ApJ...717..379B}
{Behroozi}, P.~S., {Conroy}, C., \& {Wechsler}, R.~H. 2010, \apj, 717, 379

\bibitem[{{Blanton} \& {Roweis}(2007)}]{2007AJ....133..734B}
{Blanton}, M.~R. \& {Roweis}, S. 2007, \aj, 133, 734

\bibitem[{{Brinchmann} {et~al.}(2004){Brinchmann}, {Charlot}, {White},
  {Tremonti}, {Kauffmann}, {Heckman}, \& {Brinkmann}}]{2004MNRAS.351.1151B}
{Brinchmann}, J., {Charlot}, S., {White}, S.~D.~M., {et~al.} 2004, \mnras, 351,
  1151

\bibitem[{{Cardelli} {et~al.}(1989){Cardelli}, {Clayton}, \&
  {Mathis}}]{1989ApJ...345..245C}
{Cardelli}, J.~A., {Clayton}, G.~C., \& {Mathis}, J.~S. 1989, \apj, 345, 245

\bibitem[{{Charlot} {et~al.}(2002){Charlot}, {Kauffmann}, {Longhetti},
  {Tresse}, {White}, {Maddox}, \& {Fall}}]{2002MNRAS.330..876C}
{Charlot}, S., {Kauffmann}, G., {Longhetti}, M., {et~al.} 2002, \mnras, 330,
  876

\bibitem[{{Cid Fernandes} {et~al.}(2007){Cid Fernandes}, {Asari}, {Sodr{\'e}},
  {Stasi{\'n}ska}, {Mateus}, {Torres-Papaqui}, \&
  {Schoenell}}]{2007MNRAS.375L..16C}
{Cid Fernandes}, R., {Asari}, N.~V., {Sodr{\'e}}, L., {et~al.} 2007, \mnras,
  375, L16

\bibitem[{{Cresci} {et~al.}(2019){Cresci}, {Mannucci}, \&
  {Curti}}]{2019A&A...627A..42C}
{Cresci}, G., {Mannucci}, F., \& {Curti}, M. 2019, \aap, 627, A42

\bibitem[{{Curti} {et~al.}(2017){Curti}, {Cresci}, {Mannucci}, {Marconi},
  {Maiolino}, \& {Esposito}}]{2017MNRAS.465.1384C}
{Curti}, M., {Cresci}, G., {Mannucci}, F., {et~al.} 2017, \mnras, 465, 1384

\bibitem[{{Curti} {et~al.}(2020){Curti}, {Mannucci}, {Cresci}, \&
  {Maiolino}}]{2020MNRAS.491..944C}
{Curti}, M., {Mannucci}, F., {Cresci}, G., \& {Maiolino}, R. 2020, \mnras, 491,
  944

\bibitem[{{Daddi} {et~al.}(2007){Daddi}, {Dickinson}, {Morrison}, {Chary},
  {Cimatti}, {Elbaz}, {Frayer}, {Renzini}, {Pope}, {Alexander}, {Bauer},
  {Giavalisco}, {Huynh}, {Kurk}, \& {Mignoli}}]{2007ApJ...670..156D}
{Daddi}, E., {Dickinson}, M., {Morrison}, G., {et~al.} 2007, \apj, 670, 156

\bibitem[{{Donnari} {et~al.}(2019){Donnari}, {Pillepich}, {Nelson},
  {Vogelsberger}, {Genel}, {Weinberger}, {Marinacci}, {Springel}, \&
  {Hernquist}}]{2019MNRAS.485.4817D}
{Donnari}, M., {Pillepich}, A., {Nelson}, D., {et~al.} 2019, \mnras, 485, 4817

\bibitem[{{Duarte Puertas} {et~al.}(2021){Duarte Puertas}, {Vilchez},
  {Iglesias-P{\'a}ramo}, {Drissen}, {Kehrig}, {Martin}, {P{\'e}rez-Montero}, \&
  {Arroyo-Polonio}}]{2021A&A...645A..57D}
{Duarte Puertas}, S., {Vilchez}, J.~M., {Iglesias-P{\'a}ramo}, J., {et~al.}
  2021, \aap, 645, A57

\bibitem[{{Duarte Puertas} {et~al.}(2017){Duarte Puertas}, {Vilchez},
  {Iglesias-P{\'a}ramo}, {Kehrig}, {P{\'e}rez-Montero}, \&
  {Rosales-Ortega}}]{2017A&A...599A..71D}
{Duarte Puertas}, S., {Vilchez}, J.~M., {Iglesias-P{\'a}ramo}, J., {et~al.}
  2017, \aap, 599, A71

\bibitem[{{Duarte Puertas} {et~al.}(2022){Duarte Puertas}, {Vilchez},
  {Iglesias-P{\'a}ramo}, {Moll{\'a}}, {P{\'e}rez-Montero}, {Kehrig},
  {Pilyugin}, \& {Zinchenko}}]{2022A&A...666A.186D}
{Duarte Puertas}, S., {Vilchez}, J.~M., {Iglesias-P{\'a}ramo}, J., {et~al.}
  2022, \aap, 666, A186

\bibitem[{{Duplancic} {et~al.}(2020){Duplancic}, {D{\'a}vila-Kurb{\'a}n},
  {Coldwell}, {Alonso}, \& {Galdeano}}]{2020MNRAS.493.1818D}
{Duplancic}, F., {D{\'a}vila-Kurb{\'a}n}, F., {Coldwell}, G.~V., {Alonso}, S.,
  \& {Galdeano}, D. 2020, \mnras, 493, 1818

\bibitem[{{Ellison} {et~al.}(2021){Ellison}, {Lin}, {Thorp}, {Pan}, {Scudder},
  {S{\'a}nchez}, {Bluck}, \& {Maiolino}}]{2021MNRAS.501.4777E}
{Ellison}, S.~L., {Lin}, L., {Thorp}, M.~D., {et~al.} 2021, \mnras, 501, 4777

\bibitem[{{Ellison} {et~al.}(2008){Ellison}, {Patton}, {Simard}, \&
  {McConnachie}}]{2008ApJ...672L.107E}
{Ellison}, S.~L., {Patton}, D.~R., {Simard}, L., \& {McConnachie}, A.~W. 2008,
  \apjl, 672, L107

\bibitem[{{Ellison} {et~al.}(2020){Ellison}, {Thorp}, {Pan}, {Lin}, {Scudder},
  {Bluck}, {S{\'a}nchez}, \& {Sargent}}]{2020MNRAS.492.6027E}
{Ellison}, S.~L., {Thorp}, M.~D., {Pan}, H.-A., {et~al.} 2020, \mnras, 492,
  6027

\bibitem[{{Erb} {et~al.}(2006){Erb}, {Shapley}, {Pettini}, {Steidel}, {Reddy},
  \& {Adelberger}}]{2006ApJ...644..813E}
{Erb}, D.~K., {Shapley}, A.~E., {Pettini}, M., {et~al.} 2006, \apj, 644, 813

\bibitem[{{Finlator} \& {Dav{\'e}}(2008)}]{2008MNRAS.385.2181F}
{Finlator}, K. \& {Dav{\'e}}, R. 2008, \mnras, 385, 2181

\bibitem[{{Foster} {et~al.}(2012){Foster}, {Hopkins}, {Gunawardhana},
  {Lara-L{\'o}pez}, {Sharp}, {Steele}, {Taylor}, {Driver}, {Baldry}, {Bamford},
  {Liske}, {Loveday}, {Norberg}, {Peacock}, {Alpaslan}, {Bauer},
  {Bland-Hawthorn}, {Brough}, {Cameron}, {Colless}, {Conselice}, {Croom},
  {Frenk}, {Hill}, {Jones}, {Kelvin}, {Kuijken}, {Nichol}, {Owers},
  {Parkinson}, {Pimbblet}, {Popescu}, {Prescott}, {Robotham}, {Lopez-Sanchez},
  {Sutherland}, {Thomas}, {Tuffs}, {van Kampen}, \&
  {Wijesinghe}}]{2012A&A...547A..79F}
{Foster}, C., {Hopkins}, A.~M., {Gunawardhana}, M., {et~al.} 2012, \aap, 547,
  A79

\bibitem[{{Fraser-McKelvie} {et~al.}(2022){Fraser-McKelvie}, {Cortese},
  {Groves}, {Brough}, {Bryant}, {Catinella}, {Croom}, {D'Eugenio},
  {L{\'o}pez-S{\'a}nchez}, {van de Sande}, {Sweet}, {Vaughan},
  {Bland-Hawthorn}, {Lawrence}, {Lorente}, \& {Owers}}]{2022MNRAS.510..320F}
{Fraser-McKelvie}, A., {Cortese}, L., {Groves}, B., {et~al.} 2022, \mnras, 510,
  320

\bibitem[{{Gao} \& {Solomon}(2004)}]{2004ApJ...606..271G}
{Gao}, Y. \& {Solomon}, P.~M. 2004, \apj, 606, 271

\bibitem[{{Gao} {et~al.}(2018){Gao}, {Wang}, {Kong}, {Lin}, {Liu}, {Liu},
  {Liu}, {Hu}, {Berhane Teklu}, {Chen}, \& {Zhao}}]{2018ApJ...868...89G}
{Gao}, Y., {Wang}, E., {Kong}, X., {et~al.} 2018, \apj, 868, 89

\bibitem[{{Greener} {et~al.}(2022){Greener}, {Arag{\'o}n-Salamanca},
  {Merrifield}, {Peterken}, {Sazonova}, {Haggar}, {Bizyaev}, {Brownstein},
  {Lane}, \& {Pan}}]{2022MNRAS.516.1275G}
{Greener}, M.~J., {Arag{\'o}n-Salamanca}, A., {Merrifield}, M., {et~al.} 2022,
  \mnras, 516, 1275

\bibitem[{{Heckman} {et~al.}(1990){Heckman}, {Armus}, \&
  {Miley}}]{1990ApJS...74..833H}
{Heckman}, T.~M., {Armus}, L., \& {Miley}, G.~K. 1990, \apjs, 74, 833

\bibitem[{{Hirashita} {et~al.}(2003){Hirashita}, {Buat}, \&
  {Inoue}}]{2003A&A...410...83H}
{Hirashita}, H., {Buat}, V., \& {Inoue}, A.~K. 2003, \aap, 410, 83

\bibitem[{{Hopkins} {et~al.}(2003){Hopkins}, {Miller}, {Nichol}, {Connolly},
  {Bernardi}, {G{\'o}mez}, {Goto}, {Tremonti}, {Brinkmann}, {Ivezi{\'c}}, \&
  {Lamb}}]{2003ApJ...599..971H}
{Hopkins}, A.~M., {Miller}, C.~J., {Nichol}, R.~C., {et~al.} 2003, \apj, 599,
  971

\bibitem[{Hunter(2007)}]{Hunter:2007}
Hunter, J.~D. 2007, Computing In Science \& Engineering, 9, 90

\bibitem[{{Iglesias-P{\'a}ramo} {et~al.}(2013){Iglesias-P{\'a}ramo},
  {V{\'{\i}}lchez}, {Galbany}, {S{\'a}nchez}, {Rosales-Ortega}, {Mast},
  {Garc{\'{\i}}a-Benito}, {Husemann}, {Aguerri}, {Alves}, {Bekerait{\'e}},
  {Bland-Hawthorn}, {Catal{\'a}n-Torrecilla}, {de Amorim}, {de
  Lorenzo-C{\'a}ceres}, {Ellis}, {Falc{\'o}n-Barroso}, {Flores}, {Florido},
  {Gallazzi}, {Gomes}, {Gonz{\'a}lez Delgado}, {Haines},
  {Hern{\'a}ndez-Fern{\'a}ndez}, {Kehrig}, {L{\'o}pez-S{\'a}nchez},
  {Lyubenova}, {Marino}, {Moll{\'a}}, {Monreal-Ibero}, {Mour{\~a}o},
  {Papaderos}, {Rodrigues}, {S{\'a}nchez-Bl{\'a}zquez}, {Spekkens},
  {Stanishev}, {van de Ven}, {Walcher}, {Wisotzki}, {Zibetti}, \&
  {Ziegler}}]{2013A&A...553L...7I}
{Iglesias-P{\'a}ramo}, J., {V{\'{\i}}lchez}, J.~M., {Galbany}, L., {et~al.}
  2013, \aap, 553, L7

\bibitem[{{Iglesias-P{\'a}ramo} {et~al.}(2016){Iglesias-P{\'a}ramo},
  {V{\'{\i}}lchez}, {Rosales-Ortega}, {S{\'a}nchez}, {Duarte Puertas},
  {Petropoulou}, {Gil de Paz}, {Galbany}, {Moll{\'a}},
  {Catal{\'a}n-Torrecilla}, {Castillo Morales}, {Mast}, {Husemann},
  {Garc{\'{\i}}a-Benito}, {Mendoza}, {Kehrig}, {P{\'e}rez-Montero},
  {Papaderos}, {Gomes}, {Walcher}, {Gonz{\'a}lez Delgado}, {Marino},
  {L{\'o}pez-S{\'a}nchez}, {Ziegler}, {Flores}, \&
  {Alves}}]{2016ApJ...826...71I}
{Iglesias-P{\'a}ramo}, J., {V{\'{\i}}lchez}, J.~M., {Rosales-Ortega}, F.~F.,
  {et~al.} 2016, \apj, 826, 71

\bibitem[{{Kauffmann} {et~al.}(2003{\natexlab{a}}){Kauffmann}, {Heckman},
  {Tremonti}, {Brinchmann}, {Charlot}, {White}, {Ridgway}, {Brinkmann},
  {Fukugita}, {Hall}, {Ivezi{\'c}}, {Richards}, \&
  {Schneider}}]{2003MNRAS.346.1055K}
{Kauffmann}, G., {Heckman}, T.~M., {Tremonti}, C., {et~al.} 2003{\natexlab{a}},
  \mnras, 346, 1055

\bibitem[{{Kauffmann} {et~al.}(2003{\natexlab{b}}){Kauffmann}, {Heckman},
  {White}, {Charlot}, {Tremonti}, {Brinchmann}, {Bruzual}, {Peng}, {Seibert},
  {Bernardi}, {Blanton}, {Brinkmann}, {Castander}, {Cs{\'a}bai}, {Fukugita},
  {Ivezic}, {Munn}, {Nichol}, {Padmanabhan}, {Thakar}, {Weinberg}, \&
  {York}}]{2003MNRAS.341...33K}
{Kauffmann}, G., {Heckman}, T.~M., {White}, S.~D.~M., {et~al.}
  2003{\natexlab{b}}, \mnras, 341, 33

\bibitem[{{Kennicutt}(1998)}]{1998ApJ...498..541K}
{Kennicutt}, Robert~C., J. 1998, \apj, 498, 541

\bibitem[{{Kewley} {et~al.}(2001){Kewley}, {Dopita}, {Sutherland}, {Heisler},
  \& {Trevena}}]{2001ApJ...556..121K}
{Kewley}, L.~J., {Dopita}, M.~A., {Sutherland}, R.~S., {Heisler}, C.~A., \&
  {Trevena}, J. 2001, \apj, 556, 121

\bibitem[{{Kroupa}(2001)}]{2001MNRAS.322..231K}
{Kroupa}, P. 2001, \mnras, 322, 231

\bibitem[{{Lagos} {et~al.}(2016){Lagos}, {Theuns}, {Schaye}, {Furlong},
  {Bower}, {Schaller}, {Crain}, {Trayford}, \& {Matthee}}]{2016MNRAS.459.2632L}
{Lagos}, C. d.~P., {Theuns}, T., {Schaye}, J., {et~al.} 2016, \mnras, 459, 2632

\bibitem[{{Lara-L{\'o}pez} {et~al.}(2010){Lara-L{\'o}pez}, {Cepa},
  {Bongiovanni}, {P{\'e}rez Garc{\'\i}a}, {Ederoclite}, {Casta{\~n}eda},
  {Fern{\'a}ndez Lorenzo}, {Povi{\'c}}, \&
  {S{\'a}nchez-Portal}}]{2010A&A...521L..53L}
{Lara-L{\'o}pez}, M.~A., {Cepa}, J., {Bongiovanni}, A., {et~al.} 2010, \aap,
  521, L53

\bibitem[{{Lara-L{\'o}pez} {et~al.}(2013){Lara-L{\'o}pez},
  {L{\'o}pez-S{\'a}nchez}, \& {Hopkins}}]{2013ApJ...764..178L}
{Lara-L{\'o}pez}, M.~A., {L{\'o}pez-S{\'a}nchez}, {\'A}.~R., \& {Hopkins},
  A.~M. 2013, \apj, 764, 178

\bibitem[{{Lilly} {et~al.}(2013){Lilly}, {Carollo}, {Pipino}, {Renzini}, \&
  {Peng}}]{2013ApJ...772..119L}
{Lilly}, S.~J., {Carollo}, C.~M., {Pipino}, A., {Renzini}, A., \& {Peng}, Y.
  2013, \apj, 772, 119

\bibitem[{{Lin} {et~al.}(2020){Lin}, {Ellison}, {Pan}, {Thorp}, {Su},
  {S{\'a}nchez}, {Belfiore}, {Bothwell}, {Bundy}, {Chen}, {Concas}, {Hsieh},
  {Hsieh}, {Li}, {Maiolino}, {Masters}, {Newman}, {Rowlands}, {Shi},
  {Smethurst}, {Stark}, {Xiao}, \& {Yu}}]{2020ApJ...903..145L}
{Lin}, L., {Ellison}, S.~L., {Pan}, H.-A., {et~al.} 2020, \apj, 903, 145

\bibitem[{{Lin} {et~al.}(2019){Lin}, {Pan}, {Ellison}, {Belfiore}, {Shi},
  {S{\'a}nchez}, {Hsieh}, {Rowlands}, {Ramya}, {Thorp}, {Li}, \&
  {Maiolino}}]{2019ApJ...884L..33L}
{Lin}, L., {Pan}, H.-A., {Ellison}, S.~L., {et~al.} 2019, \apjl, 884, L33

\bibitem[{{Mannucci} {et~al.}(2010){Mannucci}, {Cresci}, {Maiolino}, {Marconi},
  \& {Gnerucci}}]{2010MNRAS.408.2115M}
{Mannucci}, F., {Cresci}, G., {Maiolino}, R., {Marconi}, A., \& {Gnerucci}, A.
  2010, \mnras, 408, 2115

\bibitem[{{Mateus} {et~al.}(2006){Mateus}, {Sodr{\'e}}, {Cid Fernandes},
  {Stasi{\'n}ska}, {Schoenell}, \& {Gomes}}]{2006MNRAS.370..721M}
{Mateus}, A., {Sodr{\'e}}, L., {Cid Fernandes}, R., {et~al.} 2006, \mnras, 370,
  721

\bibitem[{{Mesa} {et~al.}(2021){Mesa}, {Alonso}, {Coldwell}, {Lambas}, \& {Nilo
  Castellon}}]{2021MNRAS.501.1046M}
{Mesa}, V., {Alonso}, S., {Coldwell}, G., {Lambas}, D.~G., \& {Nilo Castellon},
  J.~L. 2021, \mnras, 501, 1046

\bibitem[{{Namiki} {et~al.}(2021){Namiki}, {Koyama}, {Koyama}, {Yamashita},
  {Hayashi}, {Haynes}, {Shimakawa}, \& {Onodera}}]{2021ApJ...918...68N}
{Namiki}, S.~V., {Koyama}, Y., {Koyama}, S., {et~al.} 2021, \apj, 918, 68

\bibitem[{{Noeske} {et~al.}(2007){Noeske}, {Weiner}, {Faber}, {Papovich},
  {Koo}, {Somerville}, {Bundy}, {Conselice}, {Newman}, {Schiminovich}, {Le
  Floc'h}, {Coil}, {Rieke}, {Lotz}, {Primack}, {Barmby}, {Cooper}, {Davis},
  {Ellis}, {Fazio}, {Guhathakurta}, {Huang}, {Kassin}, {Martin}, {Phillips},
  {Rich}, {Small}, {Willmer}, \& {Wilson}}]{2007ApJ...660L..43N}
{Noeske}, K.~G., {Weiner}, B.~J., {Faber}, S.~M., {et~al.} 2007, \apjl, 660,
  L43

\bibitem[{{O'Donnell}(1994)}]{1994ApJ...422..158O}
{O'Donnell}, J.~E. 1994, \apj, 422, 158

\bibitem[{{O'Kane} {et~al.}(2024){O'Kane}, {Kuchner}, {Gray}, \&
  {Arag{\'o}n-Salamanca}}]{2024MNRAS.534.1682O}
{O'Kane}, C.~J., {Kuchner}, U., {Gray}, M.~E., \& {Arag{\'o}n-Salamanca}, A.
  2024, \mnras, 534, 1682

\bibitem[{{Osterbrock}(1989)}]{1989agna.book.....O}
{Osterbrock}, D.~E. 1989, {Astrophysics of gaseous nebulae and active galactic
  nuclei}

\bibitem[{pandas~development team(2020)}]{reback2020pandas}
pandas~development team, T. 2020, pandas-dev/pandas: Pandas

\bibitem[{{Pearson} {et~al.}(2021){Pearson}, {Wang}, {Brough}, {Holwerda},
  {Hopkins}, \& {Loveday}}]{2021A&A...646A.151P}
{Pearson}, W.~J., {Wang}, L., {Brough}, S., {et~al.} 2021, \aap, 646, A151

\bibitem[{{Peng} {et~al.}(2010){Peng}, {Lilly}, {Kova{\v{c}}}, {Bolzonella},
  {Pozzetti}, {Renzini}, {Zamorani}, {Ilbert}, {Knobel}, {Iovino}, {Maier},
  {Cucciati}, {Tasca}, {Carollo}, {Silverman}, {Kampczyk}, {de Ravel},
  {Sanders}, {Scoville}, {Contini}, {Mainieri}, {Scodeggio}, {Kneib}, {Le
  F{\`e}vre}, {Bardelli}, {Bongiorno}, {Caputi}, {Coppa}, {de la Torre},
  {Franzetti}, {Garilli}, {Lamareille}, {Le Borgne}, {Le Brun}, {Mignoli},
  {Perez Montero}, {Pello}, {Ricciardelli}, {Tanaka}, {Tresse}, {Vergani},
  {Welikala}, {Zucca}, {Oesch}, {Abbas}, {Barnes}, {Bordoloi}, {Bottini},
  {Cappi}, {Cassata}, {Cimatti}, {Fumana}, {Hasinger}, {Koekemoer},
  {Leauthaud}, {Maccagni}, {Marinoni}, {McCracken}, {Memeo}, {Meneux}, {Nair},
  {Porciani}, {Presotto}, \& {Scaramella}}]{2010ApJ...721..193P}
{Peng}, Y.-j., {Lilly}, S.~J., {Kova{\v{c}}}, K., {et~al.} 2010, \apj, 721, 193

\bibitem[{P\'erez \& Granger(2007)}]{PER-GRA:2007}
P\'erez, F. \& Granger, B.~E. 2007, Computing in Science and Engineering, 9, 21

\bibitem[{{P{\'e}rez-Montero}(2014)}]{2014MNRAS.441.2663P}
{P{\'e}rez-Montero}, E. 2014, \mnras, 441, 2663

\bibitem[{{P{\'e}rez-Montero} \& {Contini}(2009)}]{2009MNRAS.398..949P}
{P{\'e}rez-Montero}, E. \& {Contini}, T. 2009, \mnras, 398, 949

\bibitem[{{Pessa} {et~al.}(2021){Pessa}, {Schinnerer}, {Belfiore}, {Emsellem},
  {Leroy}, {Schruba}, {Kruijssen}, {Pan}, {Blanc}, {Sanchez-Blazquez},
  {Bigiel}, {Chevance}, {Congiu}, {Dale}, {Faesi}, {Glover}, {Grasha},
  {Groves}, {Ho}, {Jim{\'e}nez-Donaire}, {Klessen}, {Kreckel}, {Koch}, {Liu},
  {Meidt}, {Pety}, {Querejeta}, {Rosolowsky}, {Saito}, {Santoro}, {Sun},
  {Usero}, {Watkins}, \& {Williams}}]{2021A&A...650A.134P}
{Pessa}, I., {Schinnerer}, E., {Belfiore}, F., {et~al.} 2021, \aap, 650, A134

\bibitem[{{Pilyugin} \& {Grebel}(2016)}]{2016MNRAS.457.3678P}
{Pilyugin}, L.~S. \& {Grebel}, E.~K. 2016, \mnras, 457, 3678

\bibitem[{{Pilyugin} {et~al.}(2017){Pilyugin}, {Grebel}, {Zinchenko},
  {Nefedyev}, \& {Mattsson}}]{2017MNRAS.465.1358P}
{Pilyugin}, L.~S., {Grebel}, E.~K., {Zinchenko}, I.~A., {Nefedyev}, Y.~A., \&
  {Mattsson}, L. 2017, \mnras, 465, 1358

\bibitem[{{Pilyugin} {et~al.}(2013){Pilyugin}, {Lara-L{\'o}pez}, {Grebel},
  {Kehrig}, {Zinchenko}, {L{\'o}pez-S{\'a}nchez}, {V{\'{\i}}lchez}, \&
  {Mattsson}}]{2013MNRAS.432.1217P}
{Pilyugin}, L.~S., {Lara-L{\'o}pez}, M.~A., {Grebel}, E.~K., {et~al.} 2013,
  \mnras, 432, 1217

\bibitem[{{Popesso} {et~al.}(2019){Popesso}, {Concas}, {Morselli}, {Schreiber},
  {Rodighiero}, {Cresci}, {Belli}, {Erfanianfar}, {Mancini}, {Inami},
  {Dickinson}, {Ilbert}, {Pannella}, \& {Elbaz}}]{2019MNRAS.483.3213P}
{Popesso}, P., {Concas}, A., {Morselli}, L., {et~al.} 2019, \mnras, 483, 3213

\bibitem[{{Sabater} {et~al.}(2013){Sabater}, {Best}, \&
  {Argudo-Fern{\'a}ndez}}]{2013MNRAS.430..638S}
{Sabater}, J., {Best}, P.~N., \& {Argudo-Fern{\'a}ndez}, M. 2013, \mnras, 430,
  638

\bibitem[{{Salim} {et~al.}(2007){Salim}, {Rich}, {Charlot}, {Brinchmann},
  {Johnson}, {Schiminovich}, {Seibert}, {Mallery}, {Heckman}, {Forster},
  {Friedman}, {Martin}, {Morrissey}, {Neff}, {Small}, {Wyder}, {Bianchi},
  {Donas}, {Lee}, {Madore}, {Milliard}, {Szalay}, {Welsh}, \&
  {Yi}}]{2007ApJS..173..267S}
{Salim}, S., {Rich}, R.~M., {Charlot}, S., {et~al.} 2007, \apjs, 173, 267

\bibitem[{{Sampaio} {et~al.}(2024){Sampaio}, {de Carvalho},
  {Arag{\'o}n-Salamanca}, {Merrifield}, {Ferreras}, \&
  {Cornwell}}]{2024MNRAS.532..982S}
{Sampaio}, V.~M., {de Carvalho}, R.~R., {Arag{\'o}n-Salamanca}, A., {et~al.}
  2024, \mnras, 532, 982

\bibitem[{{S{\'a}nchez Almeida} \&
  {S{\'a}nchez-Menguiano}(2019)}]{2019ApJ...878L...6S}
{S{\'a}nchez Almeida}, J. \& {S{\'a}nchez-Menguiano}, L. 2019, \apjl, 878, L6

\bibitem[{{S{\'a}nchez-Menguiano} {et~al.}(2019){S{\'a}nchez-Menguiano},
  {S{\'a}nchez Almeida}, {Mu{\~n}oz-Tu{\~n}{\'o}n}, {S{\'a}nchez}, {Filho},
  {Hwang}, \& {Drory}}]{2019ApJ...882....9S}
{S{\'a}nchez-Menguiano}, L., {S{\'a}nchez Almeida}, J.,
  {Mu{\~n}oz-Tu{\~n}{\'o}n}, C., {et~al.} 2019, \apj, 882, 9

\bibitem[{{Savaglio} {et~al.}(2005){Savaglio}, {Glazebrook}, {Le Borgne},
  {Juneau}, {Abraham}, {Crampton}, {McCarthy}, {Chen}, {Marzke}, {Carlberg},
  {J{\o}gensen}, {Roth}, {Hook}, \& {Murowinski}}]{2005AIPC..761..425S}
{Savaglio}, S., {Glazebrook}, K., {Le Borgne}, D., {et~al.} 2005, in American
  Institute of Physics Conference Series, Vol. 761, The Spectral Energy
  Distributions of Gas-Rich Galaxies: Confronting Models with Data, ed. C.~C.
  {Popescu} \& R.~J. {Tuffs} (AIP), 425--428

\bibitem[{{Schlegel} {et~al.}(1998){Schlegel}, {Finkbeiner}, \&
  {Davis}}]{1998ApJ...500..525S}
{Schlegel}, D.~J., {Finkbeiner}, D.~P., \& {Davis}, M. 1998, \apj, 500, 525

\bibitem[{{Schmidt}(1959)}]{1959ApJ...129..243S}
{Schmidt}, M. 1959, \apj, 129, 243

\bibitem[{{Storey} \& {Hummer}(1995)}]{1995MNRAS.272...41S}
{Storey}, P.~J. \& {Hummer}, D.~G. 1995, \mnras, 272, 41

\bibitem[{{Stott} {et~al.}(2013){Stott}, {Sobral}, {Bower}, {Smail}, {Best},
  {Matsuda}, {Hayashi}, {Geach}, \& {Kodama}}]{2013MNRAS.436.1130S}
{Stott}, J.~P., {Sobral}, D., {Bower}, R., {et~al.} 2013, \mnras, 436, 1130

\bibitem[{{Strauss} {et~al.}(2002){Strauss}, {Weinberg}, {Lupton}, {Narayanan},
  {Annis}, {Bernardi}, {Blanton}, {Burles}, {Connolly}, {Dalcanton}, {Doi},
  {Eisenstein}, {Frieman}, {Fukugita}, {Gunn}, {Ivezi{\'c}}, {Kent}, {Kim},
  {Knapp}, {Kron}, {Munn}, {Newberg}, {Nichol}, {Okamura}, {Quinn}, {Richmond},
  {Schlegel}, {Shimasaku}, {SubbaRao}, {Szalay}, {Vanden Berk}, {Vogeley},
  {Yanny}, {Yasuda}, {York}, \& {Zehavi}}]{2002AJ....124.1810S}
{Strauss}, M.~A., {Weinberg}, D.~H., {Lupton}, R.~H., {et~al.} 2002, \aj, 124,
  1810

\bibitem[{{Tacconi} {et~al.}(2018){Tacconi}, {Genzel}, {Saintonge}, {Combes},
  {Garc{\'\i}a-Burillo}, {Neri}, {Bolatto}, {Contini}, {F{\"o}rster Schreiber},
  {Lilly}, {Lutz}, {Wuyts}, {Accurso}, {Boissier}, {Boone}, {Bouch{\'e}},
  {Bournaud}, {Burkert}, {Carollo}, {Cooper}, {Cox}, {Feruglio}, {Freundlich},
  {Herrera-Camus}, {Juneau}, {Lippa}, {Naab}, {Renzini}, {Salome}, {Sternberg},
  {Tadaki}, {{\"U}bler}, {Walter}, {Weiner}, \& {Weiss}}]{2018ApJ...853..179T}
{Tacconi}, L.~J., {Genzel}, R., {Saintonge}, A., {et~al.} 2018, \apj, 853, 179

\bibitem[{{Taylor}(2005)}]{2005ASPC..347...29T}
{Taylor}, M.~B. 2005, in Astronomical Society of the Pacific Conference Series,
  Vol. 347, Astronomical Data Analysis Software and Systems XIV, ed.
  P.~{Shopbell}, M.~{Britton}, \& R.~{Ebert}, 29

\bibitem[{{Tremonti} {et~al.}(2004){Tremonti}, {Heckman}, {Kauffmann},
  {Brinchmann}, {Charlot}, {White}, {Seibert}, {Peng}, {Schlegel}, {Uomoto},
  {Fukugita}, \& {Brinkmann}}]{2004ApJ...613..898T}
{Tremonti}, C.~A., {Heckman}, T.~M., {Kauffmann}, G., {et~al.} 2004, \apj, 613,
  898

\bibitem[{{Tumlinson} {et~al.}(2017){Tumlinson}, {Peeples}, \&
  {Werk}}]{2017ARA&A..55..389T}
{Tumlinson}, J., {Peeples}, M.~S., \& {Werk}, J.~K. 2017, \araa, 55, 389

\bibitem[{{Tumlinson} {et~al.}(2011){Tumlinson}, {Thom}, {Werk}, {Prochaska},
  {Tripp}, {Weinberg}, {Peeples}, {O'Meara}, {Oppenheimer}, {Meiring}, {Katz},
  {Dav{\'e}}, {Ford}, \& {Sembach}}]{2011Sci...334..948T}
{Tumlinson}, J., {Thom}, C., {Werk}, J.~K., {et~al.} 2011, Science, 334, 948

\bibitem[{{Vale Asari} {et~al.}(2009){Vale Asari}, {Stasi{\'n}ska}, {Cid
  Fernandes}, {Gomes}, {Schlickmann}, {Mateus}, \&
  {Schoenell}}]{2009MNRAS.396L..71V}
{Vale Asari}, N., {Stasi{\'n}ska}, G., {Cid Fernandes}, R., {et~al.} 2009,
  \mnras, 396, L71

\bibitem[{{van der Walt} {et~al.}(2011){van der Walt}, {Colbert}, \&
  {Varoquaux}}]{2011CSE....13b..22V}
{van der Walt}, S., {Colbert}, S.~C., \& {Varoquaux}, G. 2011, Computing in
  Science and Engineering, 13, 22

\bibitem[{{V{\'a}squez-Bustos} {et~al.}(2025){V{\'a}squez-Bustos},
  {Argudo-Fern{\'a}ndez}, {Boquien}, {Castillo-Baeza}, {Castillo-Rencoret}, \&
  {Ariza-Quintana}}]{2025arXiv250210078V}
{V{\'a}squez-Bustos}, P., {Argudo-Fern{\'a}ndez}, M., {Boquien}, M., {et~al.}
  2025, arXiv e-prints, arXiv:2502.10078

\bibitem[{{V{\'a}squez-Bustos} {et~al.}(2023){V{\'a}squez-Bustos},
  {Argudo-Fernandez}, {Grajales-Medina}, {Duarte Puertas}, \&
  {Verley}}]{2023A&A...670A..63V}
{V{\'a}squez-Bustos}, P., {Argudo-Fernandez}, M., {Grajales-Medina}, D.,
  {Duarte Puertas}, S., \& {Verley}, S. 2023, \aap, 670, A63

\bibitem[{{Veilleux} \& {Osterbrock}(1987)}]{1987ApJS...63..295V}
{Veilleux}, S. \& {Osterbrock}, D.~E. 1987, \apjs, 63, 295

\bibitem[{{Verley} {et~al.}(2007{\natexlab{a}}){Verley}, {Combes},
  {Verdes-Montenegro}, {Bergond}, \& {Leon}}]{2007A&A...474...43V}
{Verley}, S., {Combes}, F., {Verdes-Montenegro}, L., {Bergond}, G., \& {Leon},
  S. 2007{\natexlab{a}}, \aap, 474, 43

\bibitem[{{Verley} {et~al.}(2007{\natexlab{b}}){Verley}, {Leon},
  {Verdes-Montenegro}, {Combes}, {Sabater}, {Sulentic}, {Bergond}, {Espada},
  {Garc{\'{\i}}a}, {Lisenfeld}, \& {Odewahn}}]{2007A&A...472..121V}
{Verley}, S., {Leon}, S., {Verdes-Montenegro}, L., {et~al.} 2007{\natexlab{b}},
  \aap, 472, 121

\bibitem[{{Verley} {et~al.}(2007{\natexlab{c}}){Verley}, {Odewahn},
  {Verdes-Montenegro}, {Leon}, {Combes}, {Sulentic}, {Bergond}, {Espada},
  {Garc{\'{\i}}a}, {Lisenfeld}, \& {Sabater}}]{2007A&A...470..505V}
{Verley}, S., {Odewahn}, S.~C., {Verdes-Montenegro}, L., {et~al.}
  2007{\natexlab{c}}, \aap, 470, 505

\bibitem[{{Virtanen} {et~al.}(2020){Virtanen}, {Gommers}, {Oliphant},
  {Haberland}, {Reddy}, {Cournapeau}, {Burovski}, {Peterson}, {Weckesser},
  {Bright}, {van der Walt}, {Brett}, {Wilson}, {Jarrod Millman}, {Mayorov},
  {Nelson}, {Jones}, {Kern}, {Larson}, {Carey}, {Polat}, {Feng}, {Moore}, {Vand
  erPlas}, {Laxalde}, {Perktold}, {Cimrman}, {Henriksen}, {Quintero}, {Harris},
  {Archibald}, {Ribeiro}, {Pedregosa}, {van Mulbregt}, \&
  {Contributors}}]{2020SciPy-NMeth}
{Virtanen}, P., {Gommers}, R., {Oliphant}, T.~E., {et~al.} 2020, Nature
  Methods, 17, 261

\bibitem[{{Wang} \& {Lilly}(2021)}]{2021ApJ...910..137W}
{Wang}, E. \& {Lilly}, S.~J. 2021, \apj, 910, 137

\bibitem[{{W}es {M}c{K}inney(2010)}]{mckinney-proc-scipy-2010}
{W}es {M}c{K}inney. 2010, in {P}roceedings of the 9th {P}ython in {S}cience
  {C}onference, ed. {S}t\'efan van~der {W}alt \& {J}arrod {M}illman, 56 -- 61

\bibitem[{{Whitaker} {et~al.}(2012){Whitaker}, {van Dokkum}, {Brammer}, \&
  {Franx}}]{2012ApJ...754L..29W}
{Whitaker}, K.~E., {van Dokkum}, P.~G., {Brammer}, G., \& {Franx}, M. 2012,
  \apjl, 754, L29

\bibitem[{{Wu} \& {Zhang}(2021)}]{2021MNRAS.503.2340W}
{Wu}, Y.-Z. \& {Zhang}, W. 2021, \mnras, 503, 2340

\bibitem[{{Yates} {et~al.}(2012){Yates}, {Kauffmann}, \&
  {Guo}}]{2012MNRAS.422..215Y}
{Yates}, R.~M., {Kauffmann}, G., \& {Guo}, Q. 2012, \mnras, 422, 215

\bibitem[{{York} {et~al.}(2000){York}, {Adelman}, {Anderson}, {Anderson},
  {Annis}, {Bahcall}, {Bakken}, {Barkhouser}, {Bastian}, {Berman}, {Boroski},
  {Bracker}, {Briegel}, {Briggs}, {Brinkmann}, {Brunner}, {Burles}, {Carey},
  {Carr}, {Castander}, {Chen}, {Colestock}, {Connolly}, {Crocker}, {Csabai},
  {Czarapata}, {Davis}, {Doi}, {Dombeck}, {Eisenstein}, {Ellman}, {Elms},
  {Evans}, {Fan}, {Federwitz}, {Fiscelli}, {Friedman}, {Frieman}, {Fukugita},
  {Gillespie}, {Gunn}, {Gurbani}, {de Haas}, {Haldeman}, {Harris}, {Hayes},
  {Heckman}, {Hennessy}, {Hindsley}, {Holm}, {Holmgren}, {Huang}, {Hull},
  {Husby}, {Ichikawa}, {Ichikawa}, {Ivezi{\'c}}, {Kent}, {Kim}, {Kinney},
  {Klaene}, {Kleinman}, {Kleinman}, {Knapp}, {Korienek}, {Kron}, {Kunszt},
  {Lamb}, {Lee}, {Leger}, {Limmongkol}, {Lindenmeyer}, {Long}, {Loomis},
  {Loveday}, {Lucinio}, {Lupton}, {MacKinnon}, {Mannery}, {Mantsch}, {Margon},
  {McGehee}, {McKay}, {Meiksin}, {Merelli}, {Monet}, {Munn}, {Narayanan},
  {Nash}, {Neilsen}, {Neswold}, {Newberg}, {Nichol}, {Nicinski}, {Nonino},
  {Okada}, {Okamura}, {Ostriker}, {Owen}, {Pauls}, {Peoples}, {Peterson},
  {Petravick}, {Pier}, {Pope}, {Pordes}, {Prosapio}, {Rechenmacher}, {Quinn},
  {Richards}, {Richmond}, {Rivetta}, {Rockosi}, {Ruthmansdorfer}, {Sandford},
  {Schlegel}, {Schneider}, {Sekiguchi}, {Sergey}, {Shimasaku}, {Siegmund},
  {Smee}, {Smith}, {Snedden}, {Stone}, {Stoughton}, {Strauss}, {Stubbs},
  {SubbaRao}, {Szalay}, {Szapudi}, {Szokoly}, {Thakar}, {Tremonti}, {Tucker},
  {Uomoto}, {Vanden Berk}, {Vogeley}, {Waddell}, {Wang}, {Watanabe},
  {Weinberg}, {Yanny}, {Yasuda}, \& {SDSS Collaboration}}]{2000AJ....120.1579Y}
{York}, D.~G., {Adelman}, J., {Anderson}, Jr., J.~E., {et~al.} 2000, \aj, 120,
  1579

\bibitem[{{Zahid} {et~al.}(2017){Zahid}, {Kudritzki}, {Conroy}, {Andrews}, \&
  {Ho}}]{2017ApJ...847...18Z}
{Zahid}, H.~J., {Kudritzki}, R.-P., {Conroy}, C., {Andrews}, B., \& {Ho}, I.~T.
  2017, \apj, 847, 18

\end{thebibliography}

\end{document}